\PassOptionsToPackage{prologue,table,xcdraw}{xcolor}
\documentclass[sigconf, screen,nonacm]{acmart}

\AtBeginDocument{%
  }
\setcopyright{none}

\usepackage[]{hyperref}
\usepackage[table,xcdraw]{xcolor}
\usepackage{amsmath,amsfonts}

\usepackage{amssymb}
\usepackage{algorithm}
\usepackage{algorithmic}

\usepackage{graphicx}
\usepackage{textcomp}

\usepackage{color}
\usepackage{stfloats}
\usepackage{url}
\usepackage{verbatim}
\usepackage{listings}
\usepackage{xspace}
\usepackage{multirow}
\usepackage{amssymb}
\usepackage{booktabs}
\usepackage{float}
\usepackage{pifont}
\usepackage{ulem}
\usepackage{threeparttable}

\newcommand{\we}{\textit{SpecEE}\xspace}
\newcommand{\modified}[1]{\textcolor{black}{#1}}

\begin{document}



\title{\huge SpecEE: Accelerating Large Language Model Inference with\\Speculative Early Exiting}
\author{Jiaming Xu$^{1}$, Jiayi Pan$^{1}$, Yongkang Zhou$^{1}$, Siming Chen$^1$, Jinhao Li$^{13}$, Yaoxiu Lian$^1$, Junyi Wu$^1$, Guohao Dai$^{123\dag}$}

\affiliation{%
  \country{$^1$Shanghai Jiao Tong University, $^2$Infinigence-AI, $^3$SII} 
}

\affiliation{%
  \country{$^\dag$Corresponding author: daiguohao@sjtu.edu.cn} 
}

\begin{abstract}

Early exiting has recently emerged as a promising technique for accelerating large language models (LLMs)  by effectively reducing the hardware computation and memory access. \modified{In this paper, we identify that the LLM vocabulary serves as the runtime search space of the early exiting predictor and significantly influences the predictor workload (\textit{e.g.}, $\sim 20\%$ overall inference latency with $\sim 3\times10^4$ vocabulary size in Llama2). 
We propose a novel paradigm using speculative models to reduce this search space, while addressing three critical challenges for further predictor optimization. 
\textbf{ (1) \uline{Time-consuming} predictor with high computational complexity.} Current predictor designs leverage basic models with high-dimensional input that ignore inherent data variation and GPU parallelization opportunities, resulting in $\sim 15\%$ overall inference latency.
\textbf{(2) \uline{Under-utilization} of layer-wise predictor deployment.} }Current early exiting systems treat the predictor in each layer equally without considering the activation frequencies of layer-wise predictors, leading to $\sim 20\%$ inference overhead. 
\textbf{(3) \uline{Exponential mapping complexity} of predictor in speculative decoding.} \modified{Each token in the token tree of speculative decoding is treated  as an independent search space when applying the current early exiting mapping, leading to exponential mapping complexity and failing to incorporate the high-throughput benefits}

To address the above challenges, we present \we, a fast LLM inference engine with speculative early exiting. 
\textbf{(1) \textit{At the algorithm level}}, we 
propose the \modified{\textbf{speculation-based \uline{lightweight} predictor design}}
by exploiting the probabilistic correlation between the speculative tokens and the correct results \modified{ and high parallelism of GPUs}.
\textbf{(2) \textit{At the system level}}, we point out that not all layers need a predictor and design the \modified{\textbf{two-level \uline{heuristic} predictor scheduling engine}} based on skewed distribution and contextual similarity.
\textbf{(3) \textit{At the mapping level}}, we point out that different decoding methods share the same essential characteristics, and propose the \modified{\textbf{context-aware \uline{merged mapping} for predictor}} with efficient GPU implementations to support speculative decoding, and form a framework for various existing orthogonal acceleration techniques (\textit{e.g.}, quantization and sparse activation) on cloud and personal computer (PC) scenarios, \textbf{successfully pushing the Pareto frontier of accuracy and speedup}.
It is worth noting that \we can be applied to any LLM by negligible training overhead in advance without affecting the model's original parameters.
Extensive experiments show that \we achieves $2.25\times$ and $2.43\times$ speedup with Llama2-7B on cloud and PC scenarios respectively.
\end{abstract}

\maketitle
\pagestyle{plain}

\section{Introduction} \label{sec:intro}

\begin{figure}[t]
    \centering
    \includegraphics[width=0.48\textwidth]{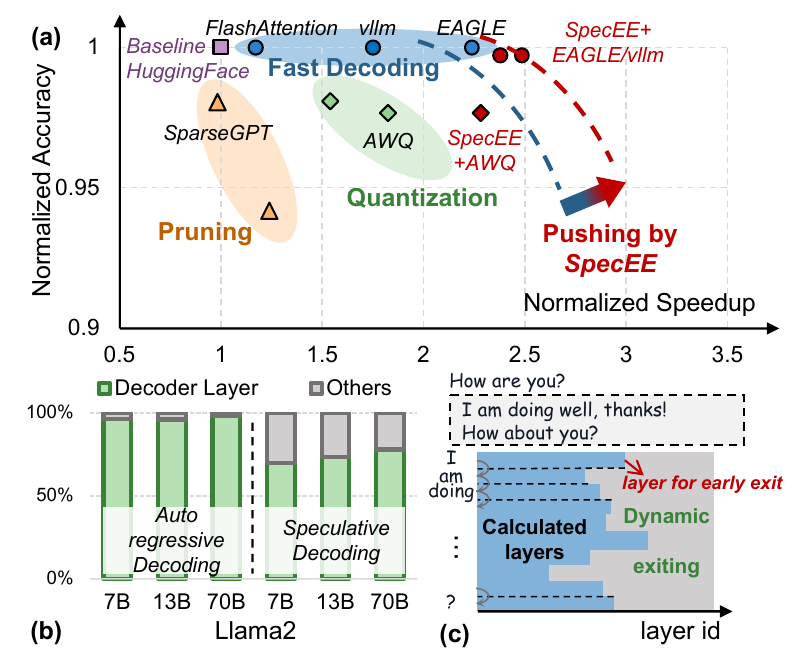}
    \vspace{-20pt}
    \caption{(a) Pareto frontier of accuracy and speedup towards LLM inference and deployment. The detailed normalized accuracy and speedup are obtained with Llama2-7B on an NVIDIA RTX 4090 GPU. (b) The ratio of the time of the decoder layer to end-to-end inference time in the original LLM. The data of two decodings is obtained based on Hugging Face~\cite{huggingface} and EAGLE~\cite{li2024eagle} frameworks. (c) Different numbers of decoder layers are needed for different token generation.}
    \vspace{-10pt}
    \label{fig:pareto}
\end{figure}

Towards the advancement of Artificial General Intelligence (AGI), generative large language models (LLMs) have been successfully applied across various domains, significantly enabling the rapid development of numerous downstream tasks (\textit{e.g.}, agent application~\cite{wu2023autogen}, code generation~\cite{code_generation_survey} and robotics~\cite{driess2023palme}). Driven by the scaling law, more and more LLMs with an increasing number of parameters (\textit{e.g.}, Grok-1~\cite{grok} with 314B parameters) have proven remarkable performance in many scenarios. However, this further results in significant memory requirements and inference latency, which poses great challenges for the deployment of practical applications.  For cloud service vendors of LLMs, the extended response time translates to increased infrastructure costs (\textit{e.g.,} energy) and suboptimal user experiences. For example, it is estimated that OpenAI consumes 260.42 MWh of energy per day~\cite{openai_consumption}, which translates into a cost of \$26,042 per day, based on U.S. industrial electricity prices of about 10 cents per kWh. This is approximately five times the average monthly income of \$4,831 in the United States~\cite{income}. 

Consequently, many previous works have explored techniques to accelerate LLM inference and reduce infrastructure cost for deployment, encompassing algorithm optimization, system enhancements, and hardware advancements~\cite{zhou2024survey,li2024largelanguagemodelinference}. Some of these works (\textit{e.g.}, fast decoding~\cite{flashattention2,deepspeed,vllm,hong2024,li2024eagle}) ensure the consistency of results, while others (\textit{e.g.}, pruning and quantization~\cite{lin2024awqactivationawareweightquantization,frantar2023sparsegpt,li2023enabling}) may lead to accuracy loss, thus forming a Pareto frontier of accuracy and speedup towards LLM inference and deployment as shown in Figure~\ref{fig:pareto}(a). However, due to the lack of consideration of the relationship between the dynamic input and the static model in these works, the multiple
cascaded layers in the original model account for \uline{$70 \sim 95\%$} of end-to-end inference shown in Figure~\ref{fig:pareto}(b), becoming the primary bottleneck for pushing the Pareto frontier forward.

\begin{figure*}[t]
    \centering
    \includegraphics[width=\textwidth]{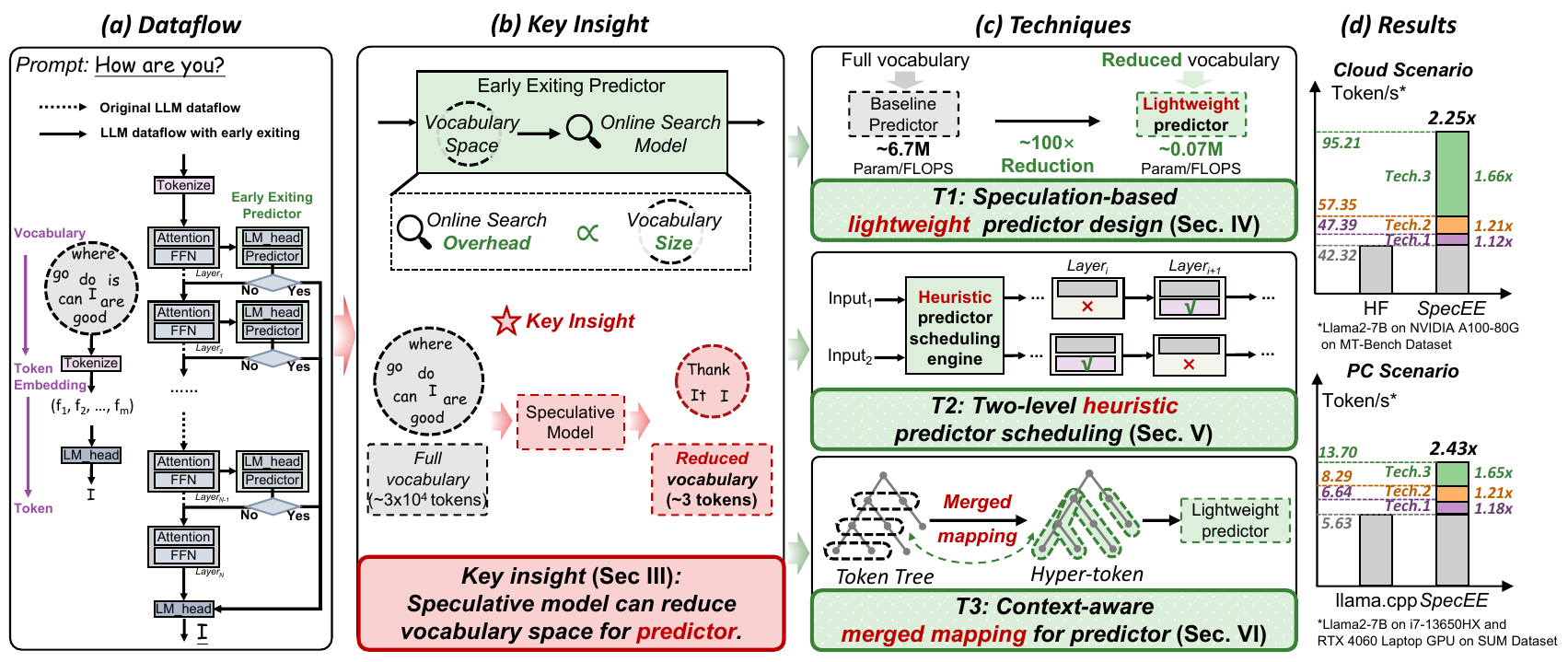}
    \vspace{-15pt}
    \caption{\modified{Overview of \we. (a) Dataflow of early exiting. (b) Key insight: Speculative model can reduce vocabulary space for predictor. (c) \uline{T}echniques on predictor optimization from Section~\ref{sec:T1} to Section~\ref{sec:T3}. (d) Results on cloud and PC scenarios.}}
    \vspace{-5pt}
    \label{fig:overview}
\end{figure*}

\modified{The inference of LLM is to generate the token with the highest probability from the full vocabulary through cascading decoder layers, which is essentially an online search problem and the search space is the full vocabulary. Early exiting algorithm is an emerging optimization in dynamic neural networks~\cite{han2021dynamicneuralnetworkssurvey,Laskaridis_2021} that aims to timely and efficiently predict when search termination occurs.} Several recent works~\cite{fan2024not,huang2024raee} have highlighted that not all decoder layers are necessary during inference in LLMs, enabling dynamic adjustments for different tokens. They suggest that the LLM parameters should be adjusted based on the complexity of the task during inference. As shown in Figure~\ref{fig:pareto}(c), during token generation,  different tokens require different forward layers to be generated. Commonly, these works entail integrating data-driven predictors (\textit{e.g.}, Support Vector Machine (SVM)~\cite{hearst1998support} and Multilayer Perceptron (MLP)~\cite{rosenblatt1958perceptron}) after each layer and structuring relevant features as input information to predict exiting. 

\modified{In this paper, we point out that the LLM vocabulary also serves as the online search space (the linear operation with the $hidden\_dim \times vocabulary\_size$ weight, called \textit{LM Head}, in LLM) of the early exiting predictors and significantly influences the workload (\textit{e.g.}, $\sim 20\%$ inference overhead with $\sim 3\times10^4$ vocabulary size of Llama2~\cite{llama} in AdaInfer~\cite{fan2024not}). Therefore, we propose \textbf{a novel paradigm using speculative models to reduce this search space by generating speculative tokens,
successfully achieving $10^4\times$ search space reduction for predictors} shown in Figure~\ref{fig:overview}(b). To apply the insight for further predictor optimization, the following challenges remain unsolved.} 


\textbf{\textit{Challenge-1:} \modified{\uline{Time-consuming} predictor with high design complexity.}} 
Current LLM early exiting predictor~\cite{fan2024not,huang2024raee} commonly need to traverse the full search space \modified{(multiplied with the complete \textit{LM Head})} to get the relevant data before prediction, \modified{and then take the raw high-dimensional ($>4\times10^3$) data as input for prediction without feature analysis and extraction. To accommodate high-dimensional input data, the predictor adopts a basic model (\textit{e.g.}, SVM in AdaInfer~\cite{fan2024not}) with high computational complexity without considering the parallelism of GPUs,}
resulting in \uline{$\sim 30\%$} overall computation and \uline{$\sim 15\%$} overall inference latency.


\textbf{\textit{Challenge-2:} \modified{\uline{Under-utilization} of layer-wise predictor deployment.}}
Current early exiting system equally treat the decoder layers of LLMs and deploy the predictor after each layer. \modified{Statistical data indicates that the success probability of the predictors follows a skewed distribution, meaning that early exiting typically occurs at a fixed set of layers for different tokens.} This implies that the computations of most other predictors are ineffective in the majority of cases,
resulting in \uline{$\sim20\%$} additional inference overhead.



\textbf{\textit{Challenge-3:} \modified{\uline{Exponential mapping complexity} of predictor in speculative decoding.}} 
\modified{Speculative decoding \cite{chen2023accelerating,cai2024medusa,li2024eagle} proposes the pattern of draft generation and token verification through tree-based token structure to address the poor throughput of autoregressive decoding. However, when applying the current early exiting mapping which aims to associate the tokens with the search space of predictors, each token of the token tree is treated as an independent seach space without considering the contextual semantics, leading to the exponential mapping complexity and the failure of incorporating the high-throughput benefits.}

To address the above challenges, we present \we, a fast LLM inference engine with speculative early exiting. The techniques of \we can be summarized into three levels as follows.

\begin{figure}[t]
    \centering
    \includegraphics[width=0.48\textwidth]{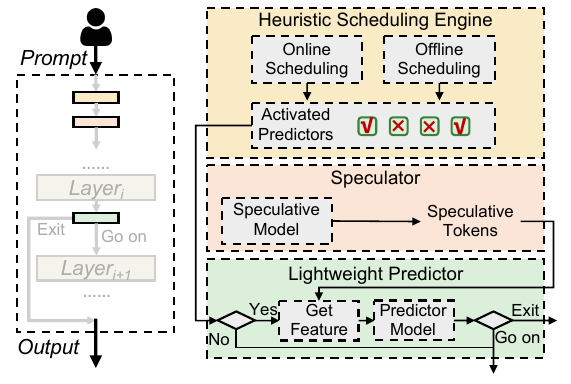}
    \vspace{-20pt}
    \caption{\modified{Architecture of \we.}}
    \vspace{-20pt}
    \label{fig:arch}
\end{figure}

\textbf{(1) Speculation-based \modified{\uline{lightweight} predictor design} at the algorithm level.}
\modified{Based on the key insight mentioned above, we point out that the probability shift of speculative tokens is strongly correlated with whether it is the correct result 
and extract the several meaningful metrics as prediction features.
To fully leverage the parallelism of GPUs, We adopt the lightweight MLP as predictor, achieving $\sim 100\times$ parameters and FLOPS reduction and $\sim 1.12\times$ end-to-end inference acceleration shown in Figure~\ref{fig:overview}(c)-T1 and (d).} 

\textbf{(2) \modified{Two-level \uline{heuristic} predictor scheduling at the system level.}}
We further point out that not all layers require predictor integration and computation based on statistical results,
and propose the two-level \modified{heuristic predictor} scheduling, which contains offline scheduling and online scheduling to achieve heuristic control over predictor integration and computation during the inference shown in Figure~\ref{fig:overview}(c)-T2. \modified{Offline scheduling allocates predictors based on skewed distribution on offline activation frequency from extensive statistical analysis}.  Online scheduling is performed runtime based on the contextual similarity of the exit layer positions, where the probability that the exit layer position of the current token is within $\pm 2$ layers of the previous five tokens exceeds $70\%$. \modified{The two-level heuristic scheduling achieves $\sim 68\%$ predictor reduction and $\sim 1.21\times$ inference acceleration shown in Figure~\ref{fig:overview}(d)}.

\textbf{(3) \modified{Context-aware \uline{merged mapping} for predictors} at the mapping level.}
Based on the contextual similarity in the exit layer positions mentioned in Technique (2), we point out that this property also applies to the tree-based speculative decoding, where contextual dependencies exist between the input token tree. Therefore, we propose \modified{the context-aware merged mapping for predictors} with efficient GPU implementations supporting speculative decoding, which merges each path in the tree-based tokens into a \textit{hyper-token} shown in Figure~\ref{fig:overview}(c)-T3, turning exponential mapping complexity into linear complexity \modified{and achieving $1.66\times$ inference acceleration shown in Figure~\ref{fig:overview}(d)}.  Moreover, due to the orthogonality,  \we also forms a framework for various existing orthogonal acceleration techniques (e.g., quantization~\cite{lin2024awqactivationawareweightquantization} and sparse activation~\cite{song2023powerinfer}) on cloud and PC scenarios, successfully pushing the Pareto frontier of accuracy and speedup shown in Figure~\ref{fig:pareto}(a).

\modified{The architecture of \we is shown in Figure~\ref{fig:arch}. After obtaining the input prompt, the heuristic scheduling engine comprising offline and online scheduling mechanisms is employed to identify the predictors that require activation. Subsequently, the speculative model is invoked to generate speculative tokens. Between each pair of consecutive decoder layers, if the predictor should be activated, features are retrieved and the predictor model is utilized to decide whether to go on with the inference process or to exit.}

We implement \we on the NVIDIA Tesla A100 80GB and RTX 4090 24GB GPUs for cloud scenario and Lenovo Legion Y7000 with i7-13650HX CPU and NVIDIA RTX 4060 Laptop 8GB GPU for PC scenario. As illustrated in Figure~\ref{fig:overview}(d), extensive experiments results on several LLMs (Llama2~\cite{llama}) show that \we achieves up to\textbf{ 2.25 $\times$ }and \textbf{2.43 $\times$} speedup compared with implementation by Hugging Face on cloud scenario and llama.cpp on PC scenario with all the techniques with negligible accuracy loss. Notably, \we can be applied to any LLM by negligible training overhead in advance without affecting the model’s original parameters.

\section{Background}

\begin{figure}[t]
    \centering
    \includegraphics[width=0.48\textwidth]{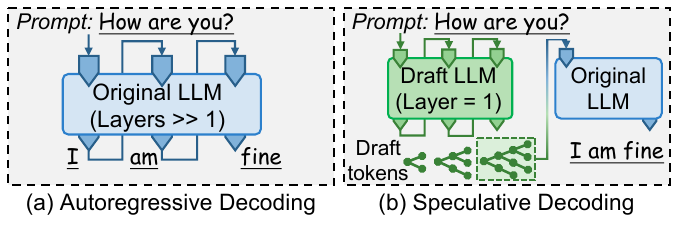}
    \vspace{-20pt}
    \caption{Two decoding methods of LLM.}
    \vspace{-10pt}
    \label{fig:decoding}
\end{figure}

\subsection{Large Language Models}

During token generation of the LLM inference, the traditional autoregressive decoding approach generates one token at a time based on input prompts and previously generated tokens, as shown in Figure~\ref{fig:decoding}(a). This ensures context - dependency for natural language processing (NLP) tasks during computation. In the attention mechanism of transformer backbone in LLMs, the generation of each token only considers the preceding content and is independent of future tokens. To reduce redundant computations, existing LLM inference systems use \textit{kv\_cache} to store keys and values of previous content. Given the self-attention mechanism's inability to handle nonlinear relationships, the Feed-Forward Network (FFN) is introduced to capture deeper, abstract features, which compensate for the limitations of self-attention mechanism.

\subsection{Speculative Decoding in LLMs}\label{sec:related:sd}


Autoregressive decoding generates a single token based on the input tokens during inference, resulting in poor throughput. Speculative decoding is proposed to address the limitation. As illustrated in Figure~\ref{fig:decoding}(b), it uses a smaller speculative draft language model (DLM) to generate speculative tokens autoregressively, forming tree-structured tokens. These tokens are then verified by the target language model (TLM) to decide which path to accept. This enables TLM to generate multiple tokens in one forward computation, achieving inference acceleration. DLM is crucial in end-to-end inference as the quality of its output determines if effective acceleration can be achieved.  As DLM has fewer parameters, methods like joint training and knowledge distillation (\textit{e.g.}, Medusa~\cite{cai2024medusa}, EAGLE~\cite{li2024eagle}) are used to align its performance with TLM.


The primitive DLM of speculative decoding often uses a smaller-scale model with the same structure.  However, it's hard to find the smallest such DLM, and different types of models can't serve as DLM for each other, limiting its versatility. LOOKAHEAD~\cite{fu2024break} ,MEDUSA~\cite{cai2024medusa} and EAGLE~\cite{li2024eagle} are highly efficient speculative decoding methods that have achieved substantial acceleration effects. However, the end-to-end time consumption of the TLM in these methods accounts for a relatively high proportion, making the TLM the main bottleneck for performance.

\begin{table}[!t]
\small
\centering
\vspace{-5pt}
\caption{Related Works on Skip Layer and Early Exiting.}
\vspace{-5pt}
\begin{tabular}{|c|c|c|c|c|}
\hline
         & Memory                       & Prediction                    & Training                     & Latency                      \\ \hline
AdaInfer~\cite{fan2024not} & \cellcolor[HTML]{DBEFDA}Low  & \cellcolor[HTML]{FAD1D1}Heavy & \cellcolor[HTML]{DBEFDA}Low  & \cellcolor[HTML]{FAD1D1}High \\ \hline
RAEE~\cite{huang2024raee}     & \cellcolor[HTML]{FAD1D1}High & \cellcolor[HTML]{FAD1D1}Heavy & \cellcolor[HTML]{DBEFDA}Low  & \cellcolor[HTML]{FAD1D1}High \\ \hline
MoD~\cite{raposo2024mixtureofdepths}      & \cellcolor[HTML]{DBEFDA}Low  & \cellcolor[HTML]{DBEFDA}Light & \cellcolor[HTML]{FAD1D1}High & \cellcolor[HTML]{DBEFDA}Low  \\ \hline
D-LLM~\cite{jiang2024dllm}    & \cellcolor[HTML]{DBEFDA}Low  & \cellcolor[HTML]{DBEFDA}Light & \cellcolor[HTML]{FAD1D1}High & \cellcolor[HTML]{DBEFDA}Low  \\ \hline
\we      & \cellcolor[HTML]{DBEFDA}Low  & \cellcolor[HTML]{DBEFDA}Light & \cellcolor[HTML]{DBEFDA}Low  & \cellcolor[HTML]{DBEFDA}Low  \\ \hline
\end{tabular}
\label{tab:works}
\vspace{-15pt}
\end{table}

\subsection{Skip Layer and Early Exiting}\label{sec:related:ee}

Several recent studies~\cite{fan2024not,huang2024raee,raposo2024mixtureofdepths,jiang2024dllm} have successfully explored the applicability of early exiting and skip layer in LLM inference. However, as shown in Table~\ref{tab:works},
existing early exiting algorithms~\cite{fan2024not,huang2024raee} introduce the prediction process with significant additional overhead in the decoding process, resulting in inefficient end-to-end inference. While existing skip layer algorithms~\cite{raposo2024mixtureofdepths,jiang2024dllm} have achieved promising performance in end-to-end inference, they require pre-training or fine-tuning of the LLM, which requires a significant cost in terms of hardware and training time.


\textbf{Skip Layer.} The Mixture-of-Depths (MoD)~\cite{raposo2024mixtureofdepths} method uses a router to let some tokens bypass blocks, and D-LLM \cite{jiang2024dllm} places a dynamic decision module before each transformer layer.  However, both MoD and D-LLM have limitations in terms of training overhead.They rely on training to learn routing or dynamic mechanisms, consuming a lot of resources and time. They often need retraining for different tasks and datasets, increasing application complexity and cost and possibly affecting their deployment and performance.

\textbf{Early Exiting.} AdaInfer~\cite{fan2024not} points out three specific features that serve as good indicators for early exiting during LLM inference. However, fetching these features needs to integrate LM head after each layer which results in deal time consumption. RAEE~\cite{huang2024raee} constructs an early exiting information database. It retrieves early exiting data based on embedding similarity and calculates the early exiting layer by probability superposition. However, its database construction is highly complex, and the inherent retrieval time leads to suboptimal end-to-end performance.

As is shown in Table~\ref{tab:works}, existing early exiting methods usually have a heavy prediction phase and high end-to-end latency, while current skip layer methods always incur high training overhead. Therefore, we aim to propose an approach that features low memory usage, light prediction, low training cost, and low latency. 

\section{\modified{Motivation}} \label{sec:motivation}

\subsection{\modified{Key Challenges of Early Exiting}}

\modified{As mentioned in Section~\ref{sec:related:ee}, AdaInfer~\cite{fan2024not} requires traversing the full vocabulary (\textit{e.g.}, $\sim 3\times 10^4$ tokens in Llama2~\cite{llama}) during prediction to obtain the probabilities of all tokens as predictor features, while RAEE~\cite{huang2024raee} requires searching the pre-built database (with a size exceeding several gigabytes) related to vocabulary , resulting in $>30\%$ overall computation and $\sim 20\%$ end-to-end inference latency.}

\modified{We analyze the online search process of the predictor and find that its overhead is primarily results from the traversal of the vocabulary, making the computational cost positively correlated with the size of the vocabulary shown in Figure~\ref{fig:overview}(b). Consequently, we identify that the vocabulary also serves as the search space for the early exiting predictor, which inherently contributes to the overhead. Therefore, we consider that the key challenge of early exiting is \textbf{how to reduce the vocabulary space using low-cost methods} involving low memory, light prediction, negligible training shown in Table~\ref{tab:works} to finally enable effective online token prediction and low end-to-end inference latency.}

\subsection{\modified{Key Insight}}

\modified{Inspired by speculation in computer system design and speculative decoding detailed in Section~\ref{sec:related:sd}, we consider that the role of DLM in speculative decoding is to generate speculative tokens for TLM. From the perspective of TLM,  the output from DLM provides a potential way to streamline the range of token selection (\textit{i.e.}, search space), even if the actual output may not always fall within this range. Furthermore, as mentioned in Section~\ref{sec:related:sd}, the goal of training DLM is to ensure that the results of TLM align as closely as possible with these speculative tokens. In other words, with a strong enough DLM, it is possible to fully limit the results of the TLM to the range of speculative tokens (\textit{i.e.}, valid small space in the insight of Figure~\ref{fig:overview}(a)).}

\modified{Therefore, we propose \textbf{a novel paradigm using speculative models to reduce the search space} shown in Figure~\ref{fig:overview}(b). The data in EAGLE~\cite{li2024eagle} shows that it only requires $\sim3\%$ memory and inference overhead of original LLM and $\sim48$ hours on RTX 3090 training overhead, which also matches our requclashirements in Table~\ref{tab:works}.}

\begin{figure*}[!t]
    \centering
    \includegraphics[width=\textwidth]{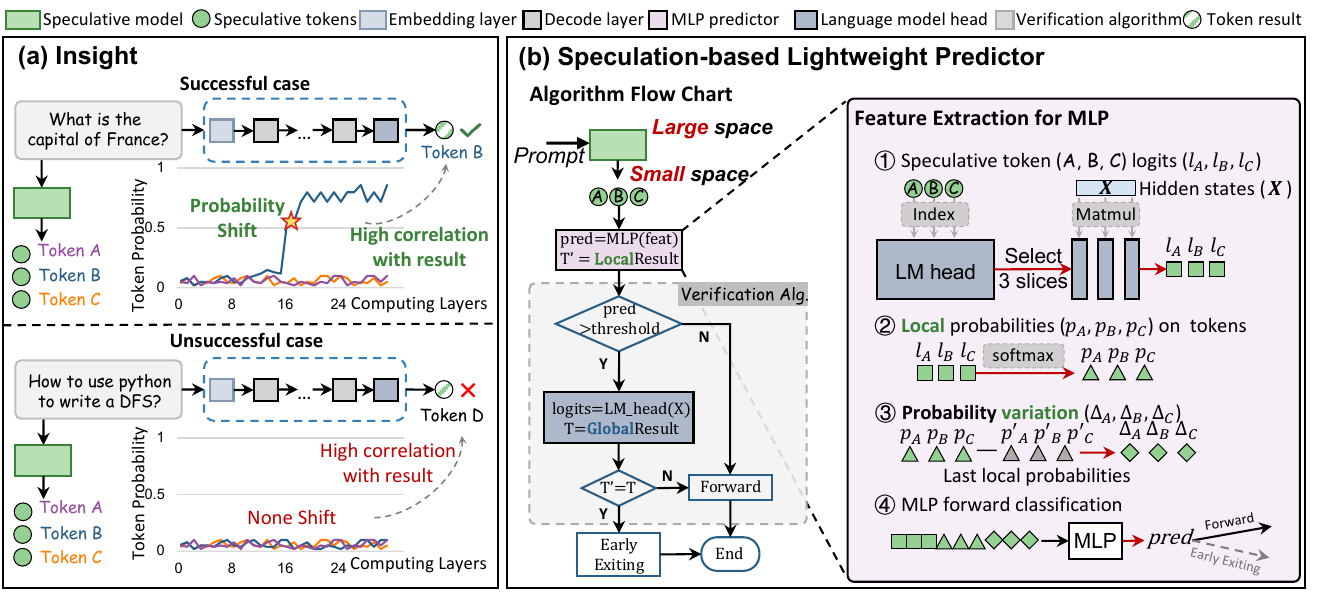}
    \vspace{-15pt}
    \caption{(a) The insight on probability shift detailed in Section~\ref{sec:T1:analysis}. (b) The algorithm flow chart and the feature extraction in speculation-based vocabulary space reduction.}
    \vspace{-10pt}
    \label{fig:t1:overview}
\end{figure*}

\begin{figure}[!t]
    \centering
    \includegraphics[width=0.48\textwidth]{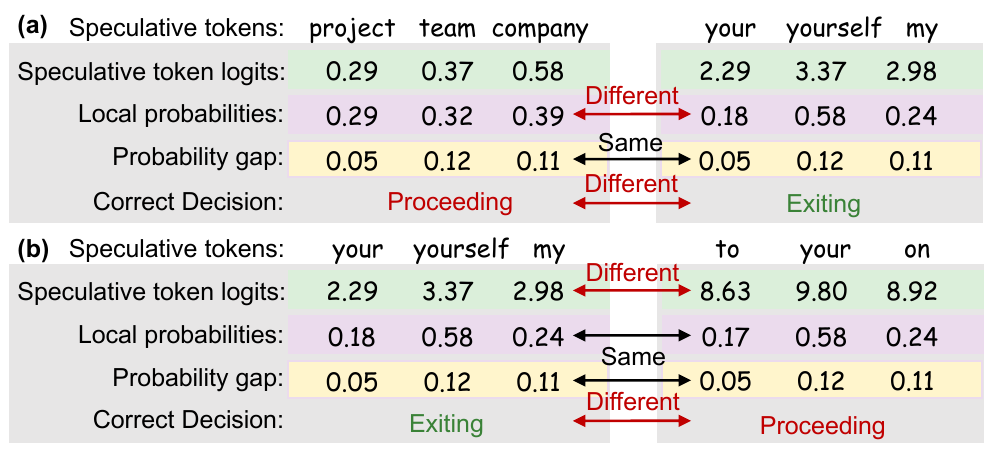}
    \vspace{-15pt}
    \caption{Analysis on feature selection. It is necessary to select three all features for prediction to prevent misjudgment.}
    \vspace{-15pt}
    \label{fig:feature}
\end{figure}

\section{\modified{Speculation-based Lightweight Predictor}} \label{sec:T1}

\subsection{\modified{Motivation: Time-consuming Predictor}}\label{sec:T1:motivation}
\modified{Though the search space can be effectively reduced by the speculative model, the design of current LLM early exiting predictor~\cite{fan2024not, huang2024raee}) still relies on directly utilizing high-dimensional raw data (\textit{e.g.}, $\sim 5\times10^3$ in Llama2-7B) retrieved from the search space as input features, without performing any feature analysis or extraction. As illustrated in Figure~\ref{fig:overview}(c)-T1, this raw high-dimensional data imposes significant demands on the predictor internal design, requiring complex architectures with a large number of parameters and computational overhead to effectively capture the implicit information contained within these high-dimensional features. Moreover, current predictor designs adopt traditional basic models (\textit{e.g.}, SVM in AdaInfer~\cite{fan2024not}) for intuitiveness and interpretability, ignoring the parallel computing opportunities provided by GPUs.}

\subsection{Insight: Probability Shift} \label{sec:T1:analysis}
We need to explore the feasibility of the new paradigm utilizing the speculative tokens generated by the speculative model as the reduced search space. As illustrated in Figure~\ref{fig:t1:overview}(a), we conduct experiments on the probability variation of tokens in the reduced space and point out that during LLM inference, if the final result token is within the reduced space, the probability of this token tends to rise sharply at a certain layer, while the probabilities of other tokens remain stable at lower values. Conversely, if the final output is not in the streamlined space, the probabilities of all tokens in the streamlined space tend to remain stable at lower values. We refer to this phenomenon as the \textbf{probability shift}.

\begin{figure*}[!t]
    \centering
    \includegraphics[width=0.98\textwidth]{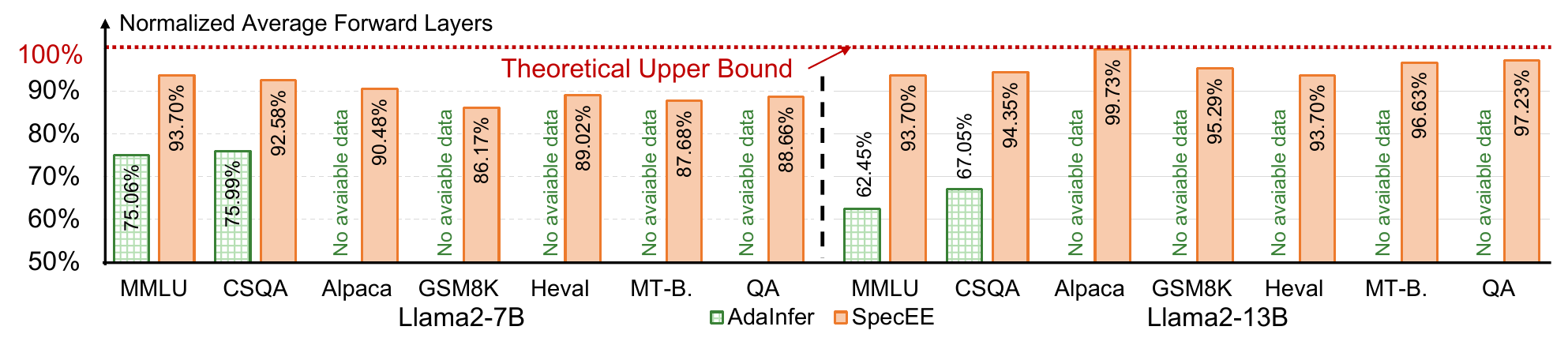}
    \vspace{-10pt}
    \caption{The gap between actual average forward layers and theoretical average forward layers of AdaInfer and \we. AdaInfer only provides the average forward layers of MMLU and CommonsenseQA in the datasets in Section~\ref{sec:exp:dataset}.}
    \vspace{-5pt}
    \label{fig:avg_layer}
\end{figure*}

\subsection{\modified{Approach: Lightweight Design}}
Based on the insight and analysis mentioned above, we design the \modified{speculation-based lightweight predictor}. The predictor design includes three parts, feature selection, judgment mechanism and correction algorithm.

\subsubsection{Feature Selection.}\label{sec:T1:Approach:Feature}
We selected speculative token logits, local probabilities, probability variation as the input features for the predictor in each layer. Below is a detailed description of each feature and the rationale behind our selection.


\textbf{(1) Speculative token logits} are the result of the matrix multiplication ($1 \times hidden\_dim \times num\_speculatives$) between the output of each layer (\textit{i.e.}, \textit{hidden\_states}) and the \textit{speculative\_lm\_head}  which refers to the columns of the \textit{lm\_head} corresponding to the speculative tokens, providing direct insight into the confidence of LLM on speculative tokens.

\textbf{(2) Local probabilities} are the result of applying the softmax function to speculative token logits. The probabilities are based on local information rather than global information, reflecting the likelihood of speculative tokens within the streamlined search space.

\textbf{(3) Probability variation} is the difference between the local probabilities in the current layer and the last layer, capturing changes in the probability across layers.

Our analysis has indicated that the probability variation of tokens is a crucial factor for prediction and we select \textbf{probability variation} as a feature. However, as illustrated in Figure~\ref{fig:feature}(a), we observe that the variation of $0.12$ can result from either $0.32 - 0.20$ or $0.58 - 0.46$. The predictor in Figure~\ref{fig:feature}(a) shouldn't allow exiting in the left while the exit probability should be higher in the right. Therefore, we consider using probability variation alone as feature is insufficient and introduce \textbf{local probabilities} as an additional feature. Moreover, the local probability may be the same when speculative token logits are different shown in Figure~\ref{fig:feature}(b). In such case, the predictor in the right conversely makes a proceeding decision and thus we further take \textbf{speculative token logits} as a feature.

\begin{figure}[!t]
    \centering
    \includegraphics[width=0.48\textwidth]{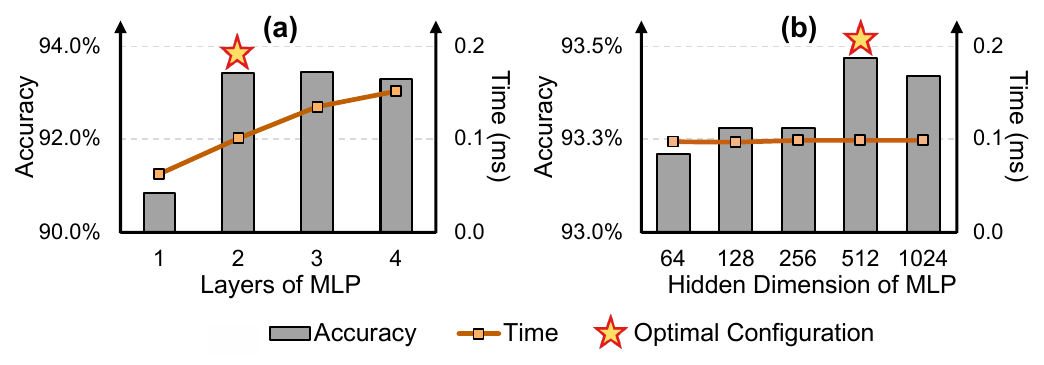}
    \vspace{-15pt}
    \caption{Design space exploration on the predictor configuration. (a) The accuracy and execution time of the predictor with changing layers and controlled hidden dimension (512). (b) The accuracy and execution time of the predictor with changing hidden dimensions and controlled layers (2).}
    \vspace{-22pt}
    \label{fig:config}
\end{figure}

\subsubsection{Judgment Mechanism.}
Based on the features mentioned above, we configure the speculative model to generate 4 speculative tokens each time, resulting in the feature dimension of 12 ($4 \times 3$). To fully leverage the high computational capacity of the GPU's Tensor Cores, we employ a two-layer MLP as the predictor with the hidden dimension of 512 instead of traditional machine learning methods (\textit{e.g.}, SVM).  The predictor employs the  ReLU activation function and sets a Sigmoid function at the output layer to handle the binary classification task. The features are fed into the predictor, and the decision to exit is determined by comparing the predictor's output to a predefined threshold (\textit{i.e.}, $0.5$).

\subsubsection{Verification Algorithm.}
As described in Section~\ref{sec:T1:Approach:Feature}, the local probabilities are derived from local information rather than global information. To verify the prediction results, we further propose the verification algorithm by incorporating global information.  As illustrated in Figure~\ref{fig:t1:overview}, we compute global token logits using the full \textit{lm\_head} and check if the token with the highest global logits is present in the speculative tokens. If it is, we exit and output that token, and if not, the model proceeds to the next layer. 



\begin{figure}[!t]
    \centering
    \includegraphics[width=0.48\textwidth]{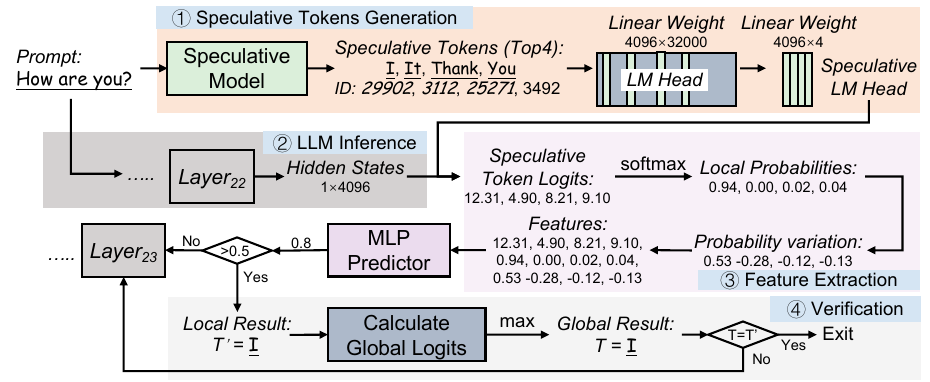}
    \vspace{-15pt}
    \caption{\modified{Example of predictor computation.}}
    \vspace{-20pt}
    \label{fig:example1}
\end{figure}

\modified{\textbf{Example.} Figure~\ref{fig:example1} shows an example of the speculation-based predictor computation. We use "How are you?" as the prompt and take the ending at layer 22 of LLM as an example. The specualtive tokens are firstly generated based on the prompt, forming the speculative LM Head. Feature extraction from the hidden states is followed during the LLM inference for the prediction Finally, the verification is performed by comparing the global token from LM Head with the local token.
}

\modified{\textbf{Design Space Exploration.} To reduce the predictor's execution time while maintaining accuracy, we focus on the number of layers and the hidden dimension. We explore the design space through using the control-variable approach shown in Figure~\ref{fig:config}. The accuracy represents the predictor's performance on the test set detailed in Section~\ref{sec:exp:predictor}. The optimal configuration is a 2-layer MLP with the hidden dimension of 512, which is our final configuration.}


\subsection{Evaluation}
The most ideal scenario for the acceleration based on early exiting is that the actual average exit layer of the method approaches the theoretical average earliest exit layer. Thus, we evaluate the average exit layer obtained by our method and the theoretical exit layer of each dataset. As illustrated in Figure~\ref{fig:avg_layer}, our method is closer to the theoretical value than AdaInfer, which is the only work about the early exiting of Llama2 models. 
Our method maintains close alignment with theoretical values across different datasets, exhibiting strong stability. 
This proximity to the theoretical exit layers is also a key reason why our approach maintains accuracy without degradation shown in Section~\ref{sec:exp:overhead:acc}.

\section{\modified{Two-level Heuristic Scheduling Engine}}
\label{sec:T2}

\subsection{Motivation} \label{sec:T2:motivation}
Based on the \modified{speculation-based lightweight predictor} proposed in Section~\ref{sec:T1}, we conducted experiments on a series of datasets in Section~\ref{sec:exp:dataset} using Llama2-7B~\cite{llama}. However, the end-to-end acceleration was not significant, showing only an average speedup of about 15\%. Despite this, the average number of executed layers was around 23, suggesting that the theoretical acceleration ratio could reach approximately $\sim 33\%$ ($32/(23+1)$). The overhead of the speculative model is roughly equivalent to the execution time of a single decoder layer. Therefore, we believe that it is the overall overhead of the predictors that slows down the end-to-end inference. The predictor overhead is defined as $T \times L$, where $T$ is the execution time of a single predictor and $L$ is the number of layers integrated with the predictor (\textit{e.g.}, $L=32$ in Llama2-7B).

\begin{figure}[!t]
    \centering
    \includegraphics[width=0.48\textwidth]{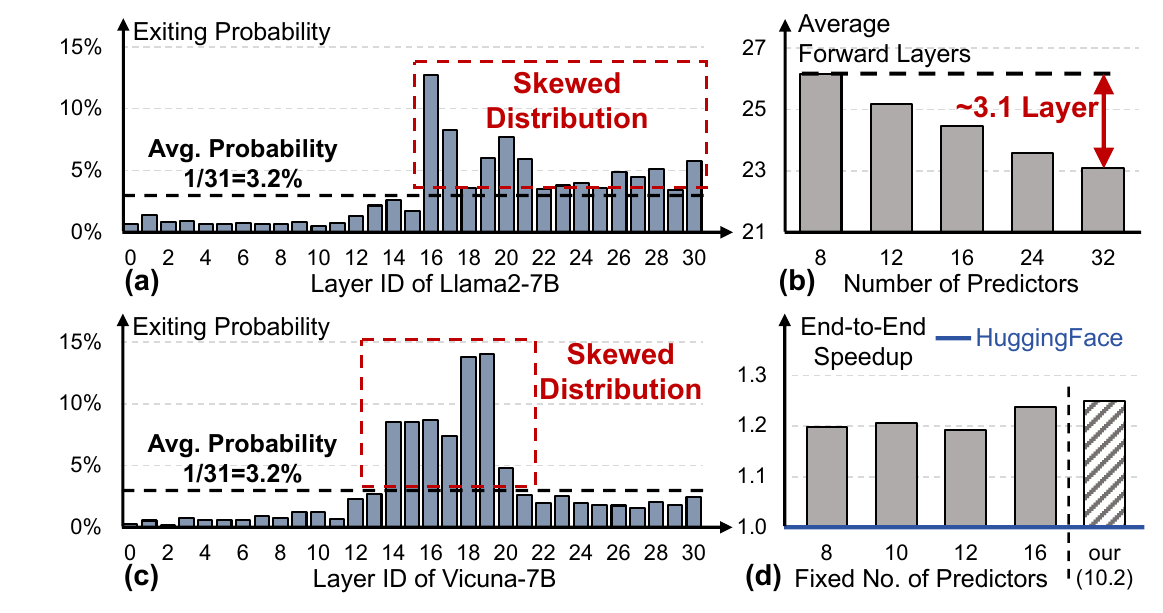}
    \vspace{-15pt}
    \caption{(a)(c) The statistical exiting probability on the 31 ($0 \sim 30$) layers in Llama2-7B and Vicuna-7B (no predictor needed for last layer). (b) The average forward layers on fixed predictors with random positions in Llama2-7B. Random positions of predictors lead to up to $\sim 3.1$ layers gap.(d) The end-to-end speedup on different fixed numbers of predictors and our dynamic predictor numbers in Llama2-7B. }
    \vspace{-15pt}
    \label{fig:distribution}
\end{figure}

Figure~\ref{fig:config} illustrates the relationship between the time overhead, accuracy, and parameter configurations of the predictor, and the final experiment is conducted with the optimal configuration (2 layers MLP with 512 hidden dimension ). Thus, reducing the overall predictor overhead can only be achieved by decreasing $L$.  Moreover, we point out that the sum of the probabilities of all layers with exit probabilities falling within the bottom $50\%$ does not exceed $20\%$ shown in Figure~\ref{fig:distribution}(a) and (c), which implies that prediction in these layers are mostly
unnecessary. However, Figure~\ref{fig:distribution}(b) indicates that blindly reducing $L$ can hinder timely exiting, leading to an increase ($\sim 3.1$ layers) in the average number of executed layers and inference latency. Therefore, we consider that the key issue is \textbf{how to accurately control the quantity ($L$) and position of predictors to achieve end-to-end inference acceleration}.

\subsection{Insights and Analysis: Skewed Distribution and Context Similarity} \label{sec:T2:analysis}

Inspired by the dynamic resource allocation in system optimization~\cite{fu2015drs,peng2018optimus,huang2014prediction}, we consider that it is necessary to dynamically adjust the number and position of predictors according to the actual situation, focusing on two key variables during LLM inference, model selection and context input. 

\textbf{Skewed Distribution.} We investigated the distribution of predictor results across two models shown in Figure~\ref{fig:distribution}(a) and (c), and identified a skewed distribution with about $50\%$ of the layers where the statistical probability of exiting is less than the average probability $3.2\%$. This skewness also varies across different models. 

\begin{figure}[!t]
    \centering
    \includegraphics[width=0.49\textwidth]{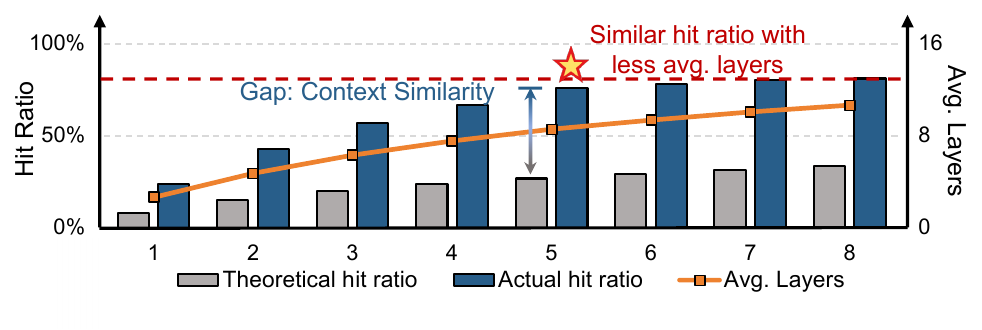}
    \vspace{-20pt}
    \caption{The explanation for context similarity. The hit ratio of the current token's exit layer within the vicinity ($\pm$2 layers) of the exit layers of the last N tokens (x-axis), as well as the average number of layers after taking the union of last N tokens' exit layers and neighboring layers.}
    \vspace{-10pt}
    \label{fig:context_simi}
\end{figure}

\textbf{Context Similarity.} Additionally, inspired by the context similarity observed in language processing~\cite{miller1991contextual} and sparse activation~\cite{mirzadeh2023relu,ma2024first}, 
\modified{we focus on the relationship of the exit layer of the current token and the last few tokens
as shown in Figure~\ref{fig:context_simi}. } Statistical results show that the exit layer of the current token has $
\sim 80\%$ probability of being near (\textit{e.g.}, $\pm 2$ layer) the exit layers of the last 5 tokens. \modified{Experiments reveal that the set consisting of the exit layers of the last 5 tokens and their neighboring layers amounts to approximately 10.2 layers on average, as shown in Figure~\ref{fig:distribution}(d).} Based on average probability calculations, the probability that the exit layer of a token falls within this set should be approximately $31.8\%$. However, experiments indicate that this probability is as high as $80\%$. Thus, we can conclude that there is a significant context similarity in the location of the exit layer.

\begin{figure}[!b]
    \centering
    \includegraphics[width=0.48\textwidth]{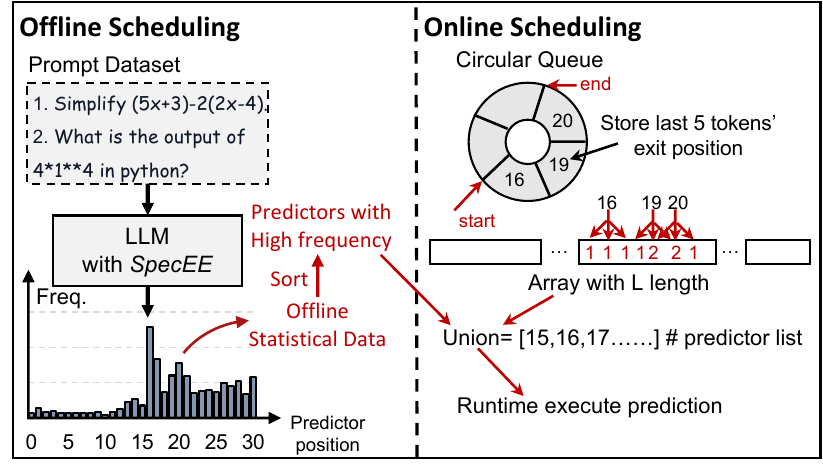}
    \vspace{-15pt}
    \caption{Dataflow of heuristic scheduling.}
    \vspace{-10pt}
    \label{fig:T2_approach}
\end{figure}

\subsection{\modified{Approach: Two-level Heuristic Scheduling}}
Based on the above insight and analysis mentioned above, we propose the two-level adaptive scheduling. The approach includes two parts, offline scheduling and online scheduling shown in Figure~\ref{fig:T2_approach}.

\textbf{Offline Scheduling.}
Given that different LLMs exhibit variations in exit probability distributions shown in Figure~\ref{fig:distribution}(a) and (c), offline scheduling is employed to collect data offline for the LLM.  It performs inference on the LLM with all predictors fully integrated using numerous prompts, collecting data from each predictor and ranking them by frequency. The result is integrated into the model as a model configuration parameter which is model-dependent and only needs to be executed offline once for a LLM.

\textbf{Online Scheduling.}
Based on the context similarity mentioned above, during the inference, we always maintain a circular queue of length $N$, representing the local context attention span (\textit{e.g.}, 5 tokens mentioned above). Additionally, we use an array with a length equal to the total number of layers ($L$). The circular queue sequentially records the exit layer positions for the last $N$ tokens, while the 
$i$-th element of the array tracks the number of times the $i$-th layer has been near (\textit{e.g.}, $\pm2$ and itself) the exit layers of last $N$ tokens recorded in the circular queue.

Finally, the quantity and position of predictors are determined by the union of a subset of results selected by the offline scheduling, and the results from the online scheduling. The performance gap between the fixed number of predictors and the dynamic number of predictors is shown in Figure~\ref{fig:distribution}(d). The dynamic selection in \we achieves the highest end-to-end speedup with fewer predictors (only $\sim 10.2$ layers).


\section{\modified{Context-aware Merged Mapping for Predictor}} \label{sec:T3}

\subsection{Motivation}

\begin{figure}[t]
    \centering
    \includegraphics[width=0.48\textwidth]{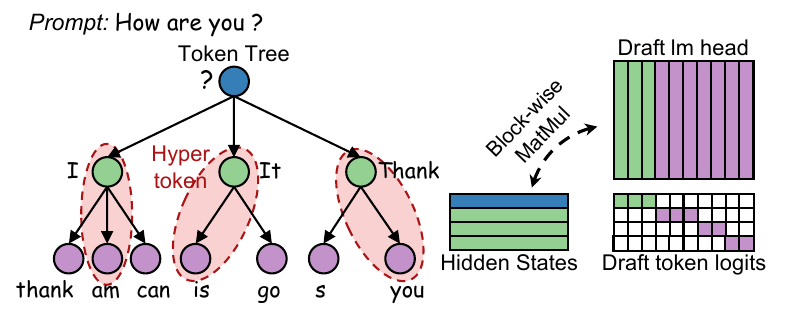}
    \vspace{-20pt}
    \caption{The hyper-token for speculative decoding and the customized GPU implementation developed based on cutlass~\cite{CUTLASS} and MegaBlocks~\cite{gale2022megablocksefficientsparsetraining} for calculating the draft token logits in \we for speculative decoding. }
    \vspace{-10pt}
    \label{fig:implementation}
\end{figure}

\modified{Speculative decoding successfully achieves high throughput through the pattern of draft generation and token verification. As illustrated in Figure~\ref{fig:implementation}, the token tree is composed of multiple tokens at each level by autoregressive generation of the speculative model. The first generation is three green tokens (Top3 probability) based on the prompt. And then these three tokens will be concatenated and fed into the speculative model and get the purple tokens at next level. All the tokens will be concatenated and fed into the target LLM to verify the tokens through one forward computation.}

When applying the early exiting during verification inference, the current mapping for predictors treats each token in the token tree as an independent search space without considering the contextual semantics. \modified{For example, the root token (\textit{?}) and its three speculative tokens (\textit{I, It, Thank}) are mapped a predictor to decide the early exiting, while the green token (\textit{I}) and its 3 speculative tokens (\textit{thank, am, can}) are also mapped a predictor at the same time. Moreover, these predictors are independent of each other, which means the overall mapping complexity is the product of the complexities of individual predictors, resulting in an exponential complexity. Therefore, we consider that the key issue is \textbf{how to design a novel mapping for speculative decoding that maintains low complexity.}}  

\subsection{\modified{Approach: Context-aware Merged Mapping}}
\textbf{Algorithm.} 
We analyze the nature of the speculative decoding and point out that early exiting shares a common essence across both decoding methods. In autoregressive decoding, early exiting is used to predict the next token based on the current token, while in speculative decoding, \modified{it is used to predict a token sequence within the token tree. However, according to the fundamental principle of early exiting, the exit position of a token sequence should be determined by the rearmost position of the exiting layers within it, reflecting an obvious Cannikin law that significantly impacts end-to-end performance. For example, if the token \textit{I} exit at the 22nd layer while the token \textit{am} exit at the 30th layer, the exiting position of the token path (\textit{I, am}) is 30th layer.} Inspired by the context similarity in Section~\ref{sec:T2:analysis},  we highlight that tokens within a token path share contextual relationships, achieving centralized exit positions and alleviating the performance loss due to Cannikin law. Thus, we propose the context-aware merge-based mapping for predictors in speculative decoding, where the tokens in a path of the token tree is merged as a single hyper-token as shown in Figure~\ref{fig:implementation}. This abstraction allows the early exiting in speculative decoding to be addressed similarly to autoregressive decoding.

\textbf{Implementation.} To efficiently compute the features of the hyper-token in Section~\ref{sec:T1:Approach:Feature} and minimize the additional overhead caused by early exiting, designed a custom GPU operator implementation shown in Figure~\ref{fig:implementation} inspired the block-wise general matrix multiplication in MegaBlocks~\cite{gale2022megablocksefficientsparsetraining} based on the group GEMM implementation of cutlass~\cite{CUTLASS}.


\subsection{Extension: Support for Orthogonal Acceleration Techniques}
\we is a dataflow developed initially based on autoregressive decoding, and it is entirely orthogonal to the techniques mentioned in Figure~\ref{fig:pareto}(a). Therefore, we have selected the following mainstream techniques in the cloud and PC scenarios for integration, successfully pushing the Pareto frontier forward. The performance is the red labels in Figure~\ref{fig:pareto}(a), and the detailed results are in Section~\ref{sec:exp}. The detailed implementation is as follows.

\begin{figure*}[!t]
    \centering
    \includegraphics[width=0.98\textwidth]{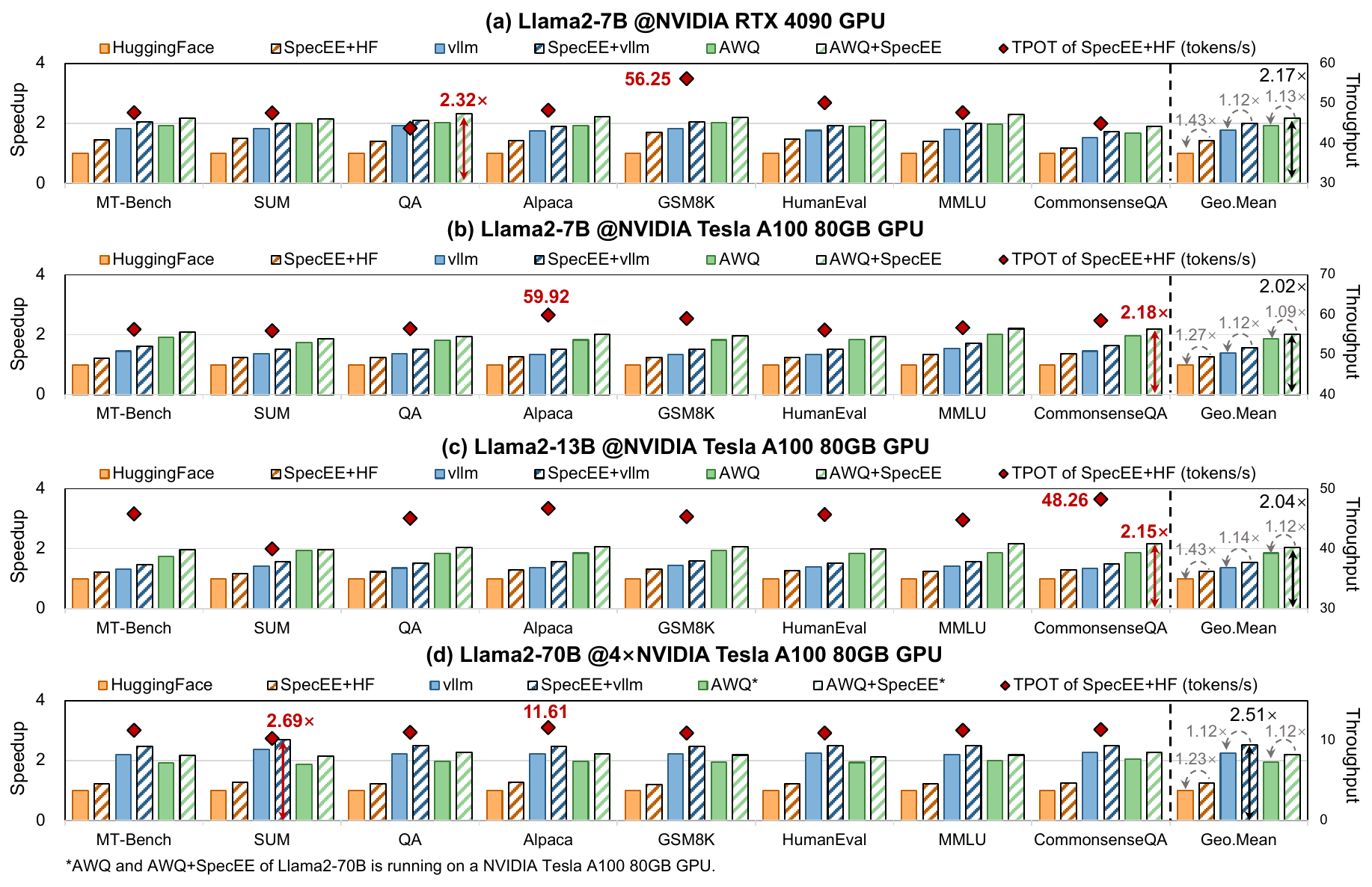}
    \vspace{-10pt}
    \caption{The speedup and throughput of Llama2-7B, Llama2-13B \modified{and Llama2-70B} on NVIDIA RTX 4090 GPU and Tesla A100 80GB GPU for autoregressive decoding in cloud scenario. }
    \vspace{-10pt}
    \label{fig:overall_cloud_auto}
\end{figure*}

\textbf{Fast Decoding.} We select the Paged Attention in vllm~\cite{vllm}  as the representative of fast decoding in the cloud scenario and implement the speculation-aided early exiting dataflow aligned with vllm including the PageAttenion usage for DLM.

\textbf{Quantization.} We select the AWQ~\cite{lin2024awqactivationawareweightquantization} with official implementation as the representative of quantization for cloud scenario and integrate the original DLM (not quantized model) for candidate token generation to achieve the early exiting dataflow.

\textbf{Sparsification.} We select the Powerinfer~\cite{song2023powerinfer} with sparse activation as the representative of sparsification for PC scenario and develop the C++ code based on the llama.cpp and the official implementation of Powerinfer, achieving the GPU-CPU hybrid inference. 



\section{Evaluation} \label{sec:exp}
\subsection{Experimental Setup}
 
We evaluate the performance of \we with various LLMs on two scenarios, cloud scenario and PC scenario. We compare the performance with several mainstream LLM inference engines in these two scenarios.


\subsubsection{Hardware Platforms}

We evaluate the performance of \we and other LLM engines on the platforms to make a comprehensive comparison. We choose two different GPUs for cloud scenarios, NVIDIA Tesla A100-80GB GPU and NVIDIA RTX 4090 24GB GPU. For the PC scenario, we select the Lenovo Legion Y7000 PC with NVIDIA RTX 4060 Laptop GPU (8GB) and Intel i7-13650HX CPU. We show the detailed configuration in Table~\ref{tab:hardware}.

\begin{table}[!h]
\centering
\caption{Hardware Platforms}
\vspace{-10pt}
\small
\begin{tabular}{@{}c|cc|c@{}}
\toprule
                                                     & \multicolumn{2}{c|}{Cloud Senario}                                                                                                                    & PC Senario                                                                        \\ \midrule
\begin{tabular}[c]{@{}c@{}}NVIDIA\\ GPU\end{tabular} & \begin{tabular}[c]{@{}c@{}}Tesla A100\\ 80GB \\ CUDA 12.1\end{tabular}         & \begin{tabular}[c]{@{}c@{}}RTX 4090\\ 24GB \\ CUDA 12.1\end{tabular} & \begin{tabular}[c]{@{}c@{}}RTX 4060 Laptop\\ 8GB \\ CUDA 12.6\end{tabular}        \\ \midrule
CPU                                                  & \begin{tabular}[c]{@{}c@{}}Intel Xeon \\ Platinum 8358 \\ 2.60GHz\end{tabular} & \begin{tabular}[c]{@{}c@{}}AMD \\ EPYC 7542\\ 2.90GHz\end{tabular}   & \begin{tabular}[c]{@{}c@{}}13th Gen Intel Core\\ i7-13650HX\\ 2.6GHz\end{tabular} \\ \bottomrule
\end{tabular}
\label{tab:hardware}
\vspace{-15pt}
\end{table}

\subsubsection{LLM Engine Baselines}

For the cloud scenario, we implement \we using the Pytorch-based front-end with the C++ and CUDA backend for NVIDIA GPUs. We integrated \we into Hugging Face~\cite{huggingface}, vllm~\cite{vllm} and AWQ~\cite{lin2024awqactivationawareweightquantization} and thus we compare the inference performance in decoding speedup and throughput with the above frameworks on two GPUs mentioned above.

For the PC scenario, we implement \we using the llama.cpp framework with the C++ and CUDA backend for NVIDIA GPUs and Intel CPUs. We integrated \we into llama.cpp~\cite{llamacpp} and PowerInfer~\cite{song2023powerinfer} and thus we compare the inference performance in decoding speedup and throughput with these two frameworks on the PC mentioned above. 
The maximum number of new tokens generated per inference is 256.

\begin{table}[!t]
\centering
\vspace{-5pt}
\caption{Model Configuration}
\vspace{-10pt}
\small
\begin{tabular}{@{}lccccc@{}}
\toprule
\multicolumn{1}{c}{Model} & Dimension & Heads  & Layers & \begin{tabular}[c]{@{}c@{}}Context \\ Length\end{tabular} \\ \midrule
Llama2-7B                 & 4096      & 32    & 32     & 4k                                                        \\
Llama2-13B                & 5120      & 40    & 40     & 4k                                                        \\
Llama2-70B                & 8192      & 64    & 80     & 4k                                                        \\
\bottomrule
\end{tabular}
\label{tab:model_conf}
\vspace{-20pt}
\end{table}

\subsubsection{Models and Datasets} \label{sec:exp:dataset}
We evaluate the performance of \we with other LLM inference engines on the chat models of Llama2-7/13/70B~\cite{llama}. Table~\ref{tab:model_conf} shows the configuration of these models. 

For evaluation on speedup and throughput, we select nine datasets in real scenarios: MT-Bench (MT-B.)~\cite{mtbench}, SUM~\cite{nallapati2016abstractive}, QA~\cite{kwiatkowski2019natural}, Alpaca~\cite{alpaca}, GSM8K~\cite{gsm8k}, HumanEval (Heval)~\cite{humaneval}, MMLU~\cite{MMLU}, CommonsenseQA (CSQA)~\cite{csqa}, and SST2~\cite{socher2013recursive}. \modified{We select seven representative datasets: MMLU, CommonsenseQA, SST2 and GSM8k for accuracy evaluation, and SUM, MT-Bench and Alpaca for perplexity (PPL) evaluation. We followed the commonly used few-shot settings in the LLM community}. 
 These tasks cover question answering, code generation, sentiment analysis, and text generation.


\begin{table*}[!h]
\vspace{-5pt}
\caption{\modified{Accuracy Evaluation}}
\vspace{-10pt}
\centering
\footnotesize
\begin{tabular}{@{}lcccccccccccccccccc@{}}
\toprule
\multicolumn{1}{l|}{\multirow{2}{*}{Task}} & \multicolumn{2}{c|}{MMLU}                & \multicolumn{2}{c|}{CommonSenseQA}       & \multicolumn{2}{c|}{SST}     & \multicolumn{2}{c|}{\modified{GSM8k}}        & \multicolumn{2}{c|}{\modified{SUM}}       & \multicolumn{2}{c|}{\modified{MT-Bench}}       & \multicolumn{2}{c}{\modified{Alpaca}}\\
\multicolumn{1}{l|}{}                      & Acc. $\uparrow$ & \multicolumn{1}{c|}{\#Avg. $L$ $\downarrow$}     & Acc. $\uparrow$ & \multicolumn{1}{c|}{\#Avg.  $L$ $\downarrow$}       & Acc. $\uparrow$ & \multicolumn{1}{c|}{\#Avg.  $L$ $\downarrow$}       & Acc. $\uparrow$ & \multicolumn{1}{c|}{\#Avg.  $L$ $\downarrow$}       & PPL $\downarrow$ & \multicolumn{1}{c|}{\#Avg.  $L$ $\downarrow$}       & PPL $\downarrow$ & \multicolumn{1}{c|}{\#Avg.  $L$ $\downarrow$}       & PPL $\downarrow$ & \#Avg. $L$  $\downarrow$  \\ \midrule
\multicolumn{15}{l}{\textbf{\textit{Llama2-7B (32 Layers)}}}                                                                                                                  \\
\multicolumn{1}{l|}{Dense}                 & 45.30 & \multicolumn{1}{c|}{32}          & 61.43 & \multicolumn{1}{c|}{32}          & 86.24 & \multicolumn{1}{c|}{32}          & \modified{20.62}   & \multicolumn{1}{c|}{\modified{32}}         & \modified{10.09}   & \multicolumn{1}{c|}{\modified{32}}         & \modified{6.49}   & \multicolumn{1}{c|}{\modified{32}}         & \modified{6.86}   & \modified{32}              \\
\multicolumn{1}{l|}{AdaInfer}              & 43.73 & \multicolumn{1}{c|}{28.91}       & 53.00 & \multicolumn{1}{c|}{27.90}       & -       &  \multicolumn{1}{c|}{-}        & \modified{0.00\textsuperscript{*}}       &  \multicolumn{1}{c|}{-}         & -   & \multicolumn{1}{c|}{-}         & -   & \multicolumn{1}{c|}{-}         & -       &  -              \\
\multicolumn{1}{l|}{\we}                   & 44.64 & \multicolumn{1}{c|}{23.16}       & 61.26 & \multicolumn{1}{c|}{22.90}       & 85.89   & \multicolumn{1}{c|}{23.55}     & \modified{20.00}   & \multicolumn{1}{c|}{\modified{23.13}}      & \modified{10.69}   & \multicolumn{1}{c|}{\modified{23.79}}      & \modified{8.44}   & \multicolumn{1}{c|}{\modified{23.22}}      & \modified{6.32}   & \modified{21.96}           \\
\multicolumn{1}{l|}{\modified{AWQ}}                 & \modified{44.61 } & \multicolumn{1}{c|}{\modified{32}}          & \modified{58.31} & \multicolumn{1}{c|}{\modified{32}}          & \modified{84.98} & \multicolumn{1}{c|}{\modified{32}}          & \modified{23.16}   & \multicolumn{1}{c|}{\modified{32}}         & \modified{7.95}   & \multicolumn{1}{c|}{\modified{32}}         & \modified{5.80}   & \multicolumn{1}{c|}{\modified{32}}         & \modified{10.01}   & \modified{32}              \\
\multicolumn{1}{l|}{\modified{AWQ+\we}}                   & \modified{44.45} & \multicolumn{1}{c|}{\modified{23.27}}       & \modified{59.05} & \multicolumn{1}{c|}{\modified{22.94}}       & \modified{84.98}   & \multicolumn{1}{c|}{\modified{22.81}}     & \modified{22.11}   & \multicolumn{1}{c|}{\modified{23.22}}      & \modified{8.08}   & \multicolumn{1}{c|}{\modified{23.50}}      & \modified{5.34}   & \multicolumn{1}{c|}{\modified{23.19}}      & \modified{5.38}   & \modified{22.28}           \\\midrule
\multicolumn{15}{l}{\textbf{\textit{Llama2-13B (40 Layers)}}}                                                                                                                 \\
\multicolumn{1}{l|}{Dense}                 & 53.58 & \multicolumn{1}{c|}{40}          & 67.57 & \multicolumn{1}{c|}{40}          & 93.00   & \multicolumn{1}{c|}{40}        & \modified{33.87}   & \multicolumn{1}{c|}{\modified{40}}         & \modified{8.76}   & \multicolumn{1}{c|}{\modified{40}}         & \modified{6.64}   & \multicolumn{1}{c|}{\modified{40}}         & \modified{4.93}   & \modified{40}              \\
\multicolumn{1}{l|}{AdaInfer}              & 52.44 & \multicolumn{1}{c|}{36.35}       & 62.48 & \multicolumn{1}{c|}{34.60}       & -       & \multicolumn{1}{c|}{-}         & -       & \multicolumn{1}{c|}{-}          & -       & \multicolumn{1}{c|}{-}          & -       & \multicolumn{1}{c|}{-}          & -       & -               \\
\multicolumn{1}{l|}{\we}                   & 53.37 & \multicolumn{1}{c|}{24.93}       & 67.16 & \multicolumn{1}{c|}{24.59}       & 92.78   & \multicolumn{1}{c|}{25.92}     & \modified{33.58}   & \multicolumn{1}{c|}{\modified{26.34}}      & \modified{7.23}   & \multicolumn{1}{c|}{\modified{27.80}}      & \modified{7.76}   & \multicolumn{1}{c|}{\modified{26.02}}      & \modified{4.82}   & \modified{24.96}           \\
\multicolumn{1}{l|}{\modified{AWQ}}                 & \modified{49.70} & \multicolumn{1}{c|}{\modified{40}}          & \modified{64.95} & \multicolumn{1}{c|}{\modified{40}}          & \modified{91.74}   & \multicolumn{1}{c|}{\modified{40}}        & \modified{28.42}   & \multicolumn{1}{c|}{\modified{40}}         & \modified{6.53}   & \multicolumn{1}{c|}{\modified{40}}         & \modified{4.66}   & \multicolumn{1}{c|}{\modified{40}}         & \modified{5.81}   & \modified{40}              \\
\multicolumn{1}{l|}{\modified{AWQ+\we}}                   & \modified{50.43} & \multicolumn{1}{c|}{\modified{28.15}}       & \modified{65.85} & \multicolumn{1}{c|}{\modified{26.57}}       & \modified{91.40}   & \multicolumn{1}{c|}{\modified{27.62}}     & \modified{27.32}   & \multicolumn{1}{c|}{\modified{28.15}}      & \modified{7.66}   & \multicolumn{1}{c|}{\modified{28.27}}      & \modified{4.00}   & \multicolumn{1}{c|}{\modified{27.22}}      & \modified{6.08}   & \modified{26.34}           \\\midrule
\multicolumn{15}{l}{\textbf{\textit{Llama2-70B (80 Layers)}}}                                                                                                                 \\
\multicolumn{1}{l|}{Dense}                 & 60.74 & \multicolumn{1}{c|}{80}          & 76.82 & \multicolumn{1}{c|}{80}          & 94.27   & \multicolumn{1}{c|}{80}        & \modified{55.79}   & \multicolumn{1}{c|}{\modified{80}}         & \modified{5.88}   & \multicolumn{1}{c|}{\modified{80}}         & \modified{4.25}   & \multicolumn{1}{c|}{\modified{80}}         & \modified{2.44}   & \modified{80}              \\
\multicolumn{1}{l|}{\we}                   & 60.54 & \multicolumn{1}{c|}{53.25}       & 76.74 & \multicolumn{1}{c|}{52.14}       & 94.04   & \multicolumn{1}{c|}{49.40}     & \modified{55.79}   & \multicolumn{1}{c|}{\modified{56.51}}      & \modified{6.07}   & \multicolumn{1}{c|}{\modified{57.58}}     & \modified{3.85}   & \multicolumn{1}{c|}{\modified{55.31}}       & \modified{1.94}   & \modified{52.88}           \\
\multicolumn{1}{l|}{\modified{AWQ}}                   & \modified{59.53} & \multicolumn{1}{c|}{\modified{80}}       & \modified{71.72} & \multicolumn{1}{c|}{\modified{80}}       & \modified{94.15}   & \multicolumn{1}{c|}{\modified{80}}     & \modified{55.05}   & \multicolumn{1}{c|}{\modified{80}}      & \modified{5.87}   & \multicolumn{1}{c|}{\modified{80}}     & \modified{4.72}   & \multicolumn{1}{c|}{\modified{80}}       & \modified{2.42}   & \modified{80}           \\
\multicolumn{1}{l|}{\modified{AWQ+\we}}                   & \modified{60.17} & \multicolumn{1}{c|}{\modified{50.14}}       & \modified{76.58} & \multicolumn{1}{c|}{\modified{53.79}}       & \modified{94.15}   & \multicolumn{1}{c|}{\modified{49.26}}     & \modified{55.08}   & \multicolumn{1}{c|}{\modified{56.34}}      & \modified{6.63}   & \multicolumn{1}{c|}{\modified{56.78}}     & \modified{4.93}   & \multicolumn{1}{c|}{\modified{52.48}}       & \modified{2.55}   & \modified{53.87}           \\\bottomrule
\end{tabular}
\begin{tablenotes}
    \footnotesize
    \item[*] $^{*}$ This data is from D-LLM\cite{jiang2024dllm}.
\end{tablenotes}
\label{tab:eva_acc}
\end{table*}

\subsection{Evaluation on Speedup and Throughput}

\begin{figure}[!t]
    \centering
    \includegraphics[width=0.48\textwidth]{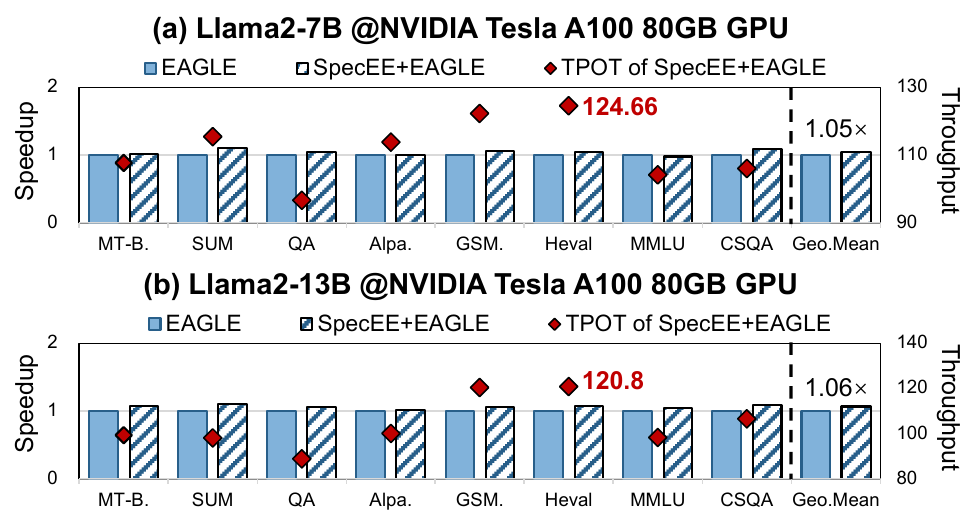}
    \vspace{-20pt}
    \caption{(a) The speedup and throughput of Llama2-7B and Llama2-13B on NVIDIA Tesla A100 80GB GPU for speculative decoding in cloud scenario.}
    \vspace{-10pt}
    \label{fig:overall_eagle}
\end{figure}

\begin{figure}[!t]
    \centering
    \includegraphics[width=0.48\textwidth]{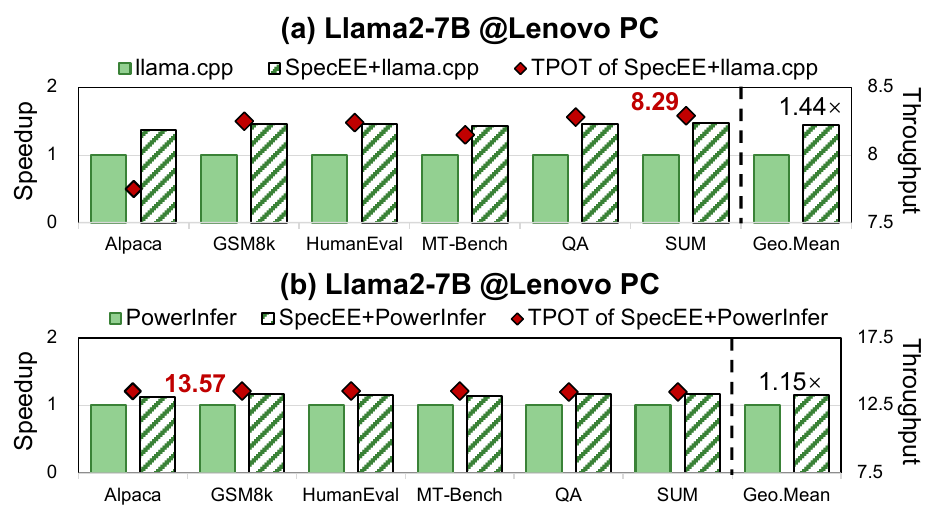}
    \vspace{-20pt}
    \caption{(a) The speedup and throughput of Llama2-7B on Lenovo PC compared with llama.cpp and PowerInfer.}
    \vspace{-15pt}
    \label{fig:overall_PC}
\end{figure}

\subsubsection{Cloud Scenario}

We divide the evaluation on speedup and throughput into two parts for autoregressive decoding and speculative decoding. 

For the autoregressive decoding, we compare the performance after the integration of \we and Hugging Face~\cite{huggingface}, vllm~\cite{vllm} and AWQ~\cite{lin2024awqactivationawareweightquantization} with the original performance of Hugging Face, vllm and AWQ using Llama2-7B, Llama2-13B \modified{and Llama2-70B} on NVIDIA GPUs with 8 datasets in Section~\ref{sec:exp:dataset}. As illustrated in Figure~\ref{fig:overall_cloud_auto}, \we achieves the average $1.43\times$, $1.12\times$ and $1.13\times$ speedup on Llama2-7B compared with Hugging Face, vllm and AWQ on RTX 4090 respectively, and achieves  average $1.27\times$, $1.12\times$ and $1.09 \times$ speedup compared with Hugging Face, vllm and AWQ on Tesla A100. And the average speedup on Tesla A100 over Hugging Face, vllm and AWQ is $1.43 \times$, $1.14\times$ and $1.12\times$ for Llama2-13B and \modified{$1.23 \times$, $1.12\times$ and $1.12\times$ for Llama2-70B}. 

For the speculative decoding, we compare the performance after the integration of \we and EAGLE~\cite{li2024eagle} with the original performance of EAGLE on Tesla A100 80GB GPU using Llama2-7B and Llama2-13B. As illustrated in Figure~\ref{fig:overall_eagle}, \we achieves the average $1.05\times$ and $1.06\times$ speedup compared with EAGLE on Llama2-7B and Llama2-13B respectively. Overall, \we achieves average $2.25\times$ speedup compared with HuggingFace on Llama2-7B.

\subsubsection{PC Scenario}

For the PC senario, we compare the performance after the integration of \we and llama.cpp~\cite{llamacpp}, PowerInfer~\cite{song2023powerinfer} with the original performance of llama.cpp and PowerInfer using Llama2-7B on Lenovo PC. As illustrated in Figure~\ref{fig:overall_PC}, \we achieves the average $1.25\times$ and $1.15\times$ speedup compared with llama.cpp and PowerInfer respectively.

\begin{figure}[!t]
    \centering
    \includegraphics[width=0.45\textwidth]{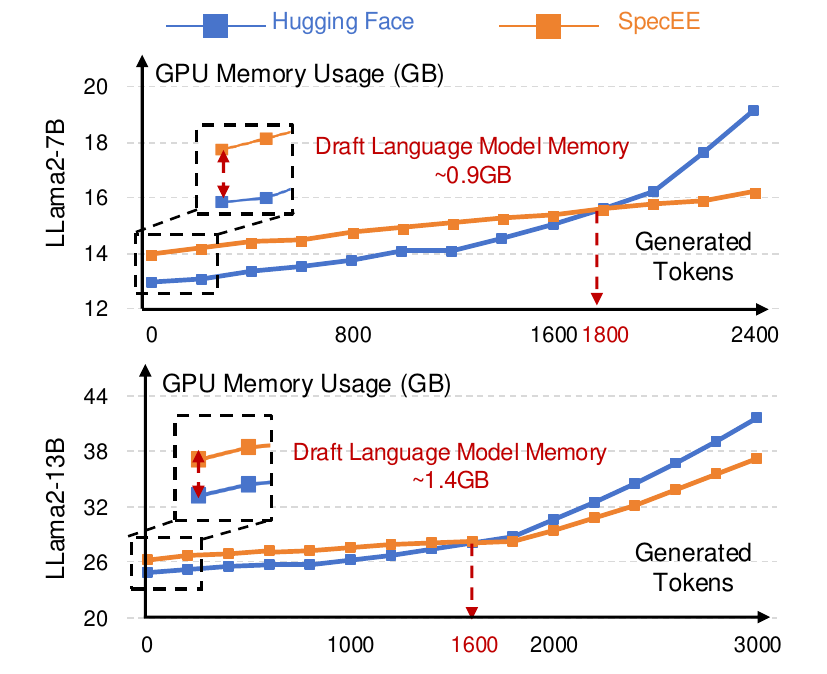}
    \vspace{-15pt}
    \caption{GPU memory usage of Llama2-7B and Llama2-13B with token generation.}
    \vspace{-15pt}
    \label{fig:memory}
\end{figure}

\subsection{\modified{Hardware Evaluation}}

\subsubsection{\modified{Energy Efficiency}}\label{sec:exp:hardware:energy}
\modified{We compare the power of dense model and \we with the Llama2-7B on NVIDIA A100 GPU (TDP 400W) using the MT-Bench dataset. We monitor GPU power consumption changes by nvidia-smi provided by NVIDIA during inference, and the statistic data shows that \we can reduce the average power from 201W to 182W, achieving $\sim 10\%$ power reduction and $\sim 1.57\times$ energy efficiency. We consider this is primarily because the predictor in \we is a memory-bound operator, and its workload is lower compared to other modules of LLM, resulting in underutilized computational resources.}

\subsubsection{\modified{Hardware Insight}} \label{sec:exp:hardware:insight}

\modified{We further profile the power and latency of the lightweight predictor on the NVIDIA A100 and Lenovo PC with RTX 4060 Laptop. \we exhibits similar latency on both the A100 and Lenovo PC but more power consumption on A100  ($\sim142W$ \textit{vs} $\sim 85W$). The A100 is an integrated training-inference architecture, while the Lenovo PC is mostly for inference. When GPUs like the A100 are used for LLM inference, power consumption should also be considered. Our advice is that future integrated training-inference GPUs could adopt a big-little core design, similar to ARM SoCs, selectively activating only a portion of CUDA cores or other computing resources to optimize power efficiency.}

\subsection{Overhead Evaluation}

\subsubsection{Accuracy Loss} \label{sec:exp:overhead:acc}

\modified{We evaluate the accuracy of the models in Section~\ref{sec:exp:dataset} with the seven datasets}. We followed the commonly used few-shot settings in the LLM community. Table~\ref{tab:eva_acc} shows that \we achieves negligible accuracy loss ($<1\%$)  compared with the original model and far outperforms the AdaInfer, whose data is obtained in its paper, both on accuracy and average forward layers.


\subsubsection{Memory Usage}
Due to the draft language model and predictors, the memory usage of \we is initially higher than original model. As illustrated in Figure~\ref{fig:memory}, the memory usage is about 0.9GB and 1.4GB more than the original model. In this paper, we choose the open-source DLM of EAGLE~\cite{li2024eagle} as the speculative model. This speculative model is the main contributor to the initial additional memory usage. As mentioned in Section~\ref{sec:T2:motivation}, the predictor is an MLP of 2 layers with 512 hidden dimension. Thus the total memory usage of all the predictors is about 416KB ($(12 \times 512 + 512 \times 1)\times32 \times 4 / 1024$) in Llama2-7B with 4 draft tokens. Compared to the DLM, the memory usage of predictors can be negligible.

\subsubsection{Speculative Model Training}

The model training overhead of \we is very low compared to the works on the skip layer described in Section~\ref{sec:related:ee}. \we only needs to train a speculative model and the preditors. In this paper, we select the DLM provided by EAGLE~\cite{li2024eagle} as the speculative model. As described in the paper of EAGLE, the speculative model for Llama2-7B only needs 24 hours of training using an RTX 3090 GPU. 

\begin{figure}[!t]
    \centering
    \includegraphics[width=0.49\textwidth]{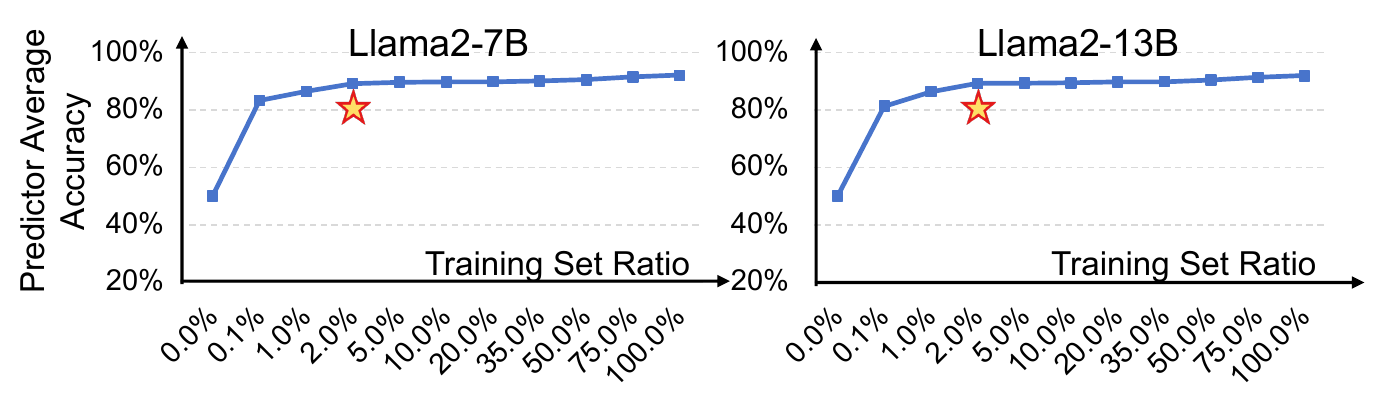}
    \vspace{-15pt}
    \caption{\modified{Predictor training of Llama2-7B/13B.}}
    \vspace{-15pt}
    \label{fig:predictor_train}
\end{figure}

\subsubsection{\modified{Predictor Offline Training and Runtime Overhead}} \label{sec:exp:predictor}
\modified{We use the MT-Bench dataset as the prompt for inference and obtain the intermediate layer (\textit{e.g.}, $0\sim30$ in Llama2-7B without the last layer) features as training data, along with the token generated by early exiting at the intermediate layer. This token is compared with the token generated after all layers, and the label is set to True if they match, or False otherwise. We totally get about 16K training data for each predictor, which takes about 1 hour on NVIDIA A100 80GB GPU. And then the training of all predictors takes about 10 minutes with total data. Figure~\ref{fig:predictor_train} shows relationship between the training data size and predictor accuracy. We only need about $\sim2\%$ training data to achieve good performance, which totally only takes about 5 minutes. We also profile the runtime overhead of predictor in Llama2-7B on a NVIDIA A100 GPU. The inference of \we is  $\sim0.016$s/token while the overhead of predictors is $0.0009$s/token, which is about $5.6\%$ inference latency.}

\subsection{Ablation Study}
We select the Llama2-7B on NVIDIA Tesla A100-80GB for ablation study and select the Hugging Face as the code base and the baseline. The overall results are in Figure~\ref{fig:ablation}.

\subsubsection{T1: Speculation-based vocabulary space reduction}
The speculation-based vocabulary space reduction aims to reduce the vocabulary space through the speculative model. For the Llama2-7B on NVIDIA Tesla A100-80GB GPU, \we only achieves about $1.08\times$ speedup across the 8 datasets shown in Figure~\ref{fig:ablation}. We have analyzed that the inefficiency is caused by redundant predictor integration and computation in Section~\ref{sec:T2:motivation}.

\subsubsection{T2: Two-level adaptive scheduling for efficient token prediction}
Two-level adaptive scheduling is thus proposed for efficient token prediction. Integrated with this technique, \we finally achieves an average $1.27 \times$ speedup across the 8 datasets for the Llama2-7B on NVIDIA Tesla A100-80GB GPU shown in Figure~\ref{fig:ablation}.

\subsubsection{T3: Merge-based multi-vocabulary to hyper-token mapping}
\begin{figure}[!t]
    \centering
    \includegraphics[width=0.48\textwidth]{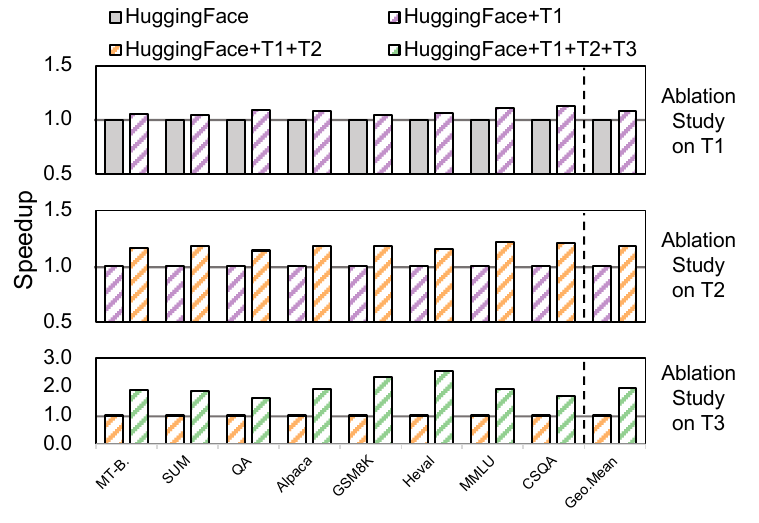}
    \vspace{-20pt}
    \caption{Ablation study of three techniques in \we.}
    \vspace{-10pt}
    \label{fig:ablation}
\end{figure}

Figure\ref{fig:ablation} shows the overall performance of \we with techniques 1 to 3, achieving the outstanding performance. For a fair comparison, we have compared the \we with EAGLE shown in Figure~\ref{fig:overall_eagle}. \we achieves about $1.06\times$ compared with EAGLE~\cite{li2024eagle}.

\section{Conclusion} \label{sec:conclusion}

\modified{In this paper, we propose the novel paradigm using the speculative model to reduce the search space, providing a new perspective to consider LLM acceleration. We think the methodology and perspective can be extended to further
studies on machine learning architecture and system design considering search space reduction.}

\modified{We apply the paradigm to the early exiting for LLM acceleration, and present the \we, a fast LLM inference engine with speculative early exiting. \we proposes three techniques for further predictor optimization at three levels of algorithm, system and mapping, }and achieves $2.25\times$ and $2.43\times$ speedup on Llama2-7B in cloud and PC scenarios, respectively, successfully pushing the Pareto frontier of accuracy and speedup.

\section*{Acknowledgment}
This work was sponsored by the National Natural Science Foundation of China (No. 62104128, U21B2031), Shanghai Rising-Star Program (No. 24QB2706200) and Beijing Douyin Information Service Co., Ltd.



\bibliographystyle{plain}
\bibliography{main}

\begin{thebibliography}{10}

\bibitem{deepspeed}
Reza~Yazdani Aminabadi, Samyam Rajbhandari, Ammar~Ahmad Awan, Cheng Li, Du~Li, Elton Zheng, Olatunji Ruwase, Shaden Smith, Minjia Zhang, Jeff Rasley, et~al.
\newblock Deepspeed-inference: enabling efficient inference of transformer models at unprecedented scale.
\newblock In {\em SC22: International Conference for High Performance Computing, Networking, Storage and Analysis}, pages 1--15. IEEE, 2022.

\bibitem{cai2024medusa}
Tianle Cai, Yuhong Li, Zhengyang Geng, Hongwu Peng, Jason~D. Lee, Deming Chen, and Tri Dao.
\newblock Medusa: Simple llm inference acceleration framework with multiple decoding heads, 2024.

\bibitem{income}
CEIC.
\newblock United states monthly earnings.
\newblock [Online], 2024.
\newblock \url{https://www.ceicdata.com/en/indicator/united-states/monthly-earnings}.

\bibitem{chen2023accelerating}
Charlie Chen, Sebastian Borgeaud, Geoffrey Irving, Jean-Baptiste Lespiau, Laurent Sifre, and John Jumper.
\newblock Accelerating large language model decoding with speculative sampling.
\newblock {\em arXiv preprint arXiv:2302.01318}, 2023.

\bibitem{humaneval}
Mark Chen, Jerry Tworek, Heewoo Jun, Qiming Yuan, Henrique~Ponde de~Oliveira~Pinto, Jared Kaplan, Harri Edwards, Yuri Burda, Nicholas Joseph, Greg Brockman, Alex Ray, Raul Puri, Gretchen Krueger, Michael Petrov, Heidy Khlaaf, Girish Sastry, Pamela Mishkin, Brooke Chan, Scott Gray, Nick Ryder, Mikhail Pavlov, Alethea Power, Lukasz Kaiser, Mohammad Bavarian, Clemens Winter, Philippe Tillet, Felipe~Petroski Such, Dave Cummings, Matthias Plappert, Fotios Chantzis, Elizabeth Barnes, Ariel Herbert-Voss, William~Hebgen Guss, Alex Nichol, Alex Paino, Nikolas Tezak, Jie Tang, Igor Babuschkin, Suchir Balaji, Shantanu Jain, William Saunders, Christopher Hesse, Andrew~N. Carr, Jan Leike, Josh Achiam, Vedant Misra, Evan Morikawa, Alec Radford, Matthew Knight, Miles Brundage, Mira Murati, Katie Mayer, Peter Welinder, Bob McGrew, Dario Amodei, Sam McCandlish, Ilya Sutskever, and Wojciech Zaremba.
\newblock Evaluating large language models trained on code, 2021.

\bibitem{gsm8k}
Karl Cobbe, Vineet Kosaraju, Mohammad Bavarian, Mark Chen, Heewoo Jun, Lukasz Kaiser, Matthias Plappert, Jerry Tworek, Jacob Hilton, Reiichiro Nakano, Christopher Hesse, and John Schulman.
\newblock Training verifiers to solve math word problems, 2021.

\bibitem{flashattention2}
Tri Dao.
\newblock Flashattention-2: Faster attention with better parallelism and work partitioning.
\newblock {\em arXiv preprint arXiv:2307.08691}, 2023.

\bibitem{driess2023palme}
Danny Driess, Fei Xia, Mehdi S.~M. Sajjadi, Corey Lynch, Aakanksha Chowdhery, Brian Ichter, Ayzaan Wahid, Jonathan Tompson, Quan Vuong, Tianhe Yu, Wenlong Huang, Yevgen Chebotar, Pierre Sermanet, Daniel Duckworth, Sergey Levine, Vincent Vanhoucke, Karol Hausman, Marc Toussaint, Klaus Greff, Andy Zeng, Igor Mordatch, and Pete Florence.
\newblock Palm-e: An embodied multimodal language model.
\newblock In {\em arXiv preprint arXiv:2303.03378}, 2023.

\bibitem{fan2024not}
Siqi Fan, Xin Jiang, Xiang Li, Xuying Meng, Peng Han, Shuo Shang, Aixin Sun, Yequan Wang, and Zhongyuan Wang.
\newblock Not all layers of llms are necessary during inference.
\newblock {\em arXiv preprint arXiv:2403.02181}, 2024.

\bibitem{frantar2023sparsegpt}
Elias Frantar and Dan Alistarh.
\newblock Sparsegpt: Massive language models can be accurately pruned in one-shot.
\newblock In {\em International Conference on Machine Learning}, pages 10323--10337. PMLR, 2023.

\bibitem{fu2015drs}
Tom~ZJ Fu, Jianbing Ding, Richard~TB Ma, Marianne Winslett, Yin Yang, and Zhenjie Zhang.
\newblock Drs: Dynamic resource scheduling for real-time analytics over fast streams.
\newblock In {\em 2015 IEEE 35th International Conference on Distributed Computing Systems}, pages 411--420. IEEE, 2015.

\bibitem{fu2024break}
Yichao Fu, Peter Bailis, Ion Stoica, and Hao Zhang.
\newblock Break the sequential dependency of llm inference using lookahead decoding, 2024.

\bibitem{gale2022megablocksefficientsparsetraining}
Trevor Gale, Deepak Narayanan, Cliff Young, and Matei Zaharia.
\newblock Megablocks: Efficient sparse training with mixture-of-experts, 2022.

\bibitem{llamacpp}
Georgi Gerganov.
\newblock Llm inference in c/c++.
\newblock [Online], 2023.
\newblock \url{https://github.com/ggerganov/llama.cpp}.

\bibitem{han2021dynamicneuralnetworkssurvey}
Yizeng Han, Gao Huang, Shiji Song, Le~Yang, Honghui Wang, and Yulin Wang.
\newblock Dynamic neural networks: A survey, 2021.

\bibitem{hearst1998support}
Marti~A. Hearst, Susan~T Dumais, Edgar Osuna, John Platt, and Bernhard Scholkopf.
\newblock Support vector machines.
\newblock {\em IEEE Intelligent Systems and their applications}, 13(4):18--28, 1998.

\bibitem{MMLU}
Dan Hendrycks, Collin Burns, Steven Basart, Andy Zou, Mantas Mazeika, Dawn Song, and Jacob Steinhardt.
\newblock Measuring massive multitask language understanding, 2021.

\bibitem{hong2024}
Ke~Hong, Guohao Dai, Jiaming Xu, Qiuli Mao, Xiuhong Li, Jun Liu, Yuhan Dong, Yu~Wang, et~al.
\newblock Flashdecoding++: Faster large language model inference with asynchronization, flat gemm optimization, and heuristics.
\newblock {\em Proceedings of Machine Learning and Systems}, 6:148--161, 2024.

\bibitem{huang2024raee}
Lianming Huang, Shangyu Wu, Yufei Cui, Ying Xiong, Xue Liu, Tei-Wei Kuo, Nan Guan, and Chun~Jason Xue.
\newblock Raee: A training-free retrieval-augmented early exiting framework for efficient inference.
\newblock {\em arXiv preprint arXiv:2405.15198}, 2024.

\bibitem{huang2014prediction}
Qingjia Huang, Kai Shuang, Peng Xu, Jian Li, Xu~Liu, and Sen Su.
\newblock Prediction-based dynamic resource scheduling for virtualized cloud systems.
\newblock {\em Journal of Networks}, 9(2):375, 2014.

\bibitem{code_generation_survey}
Juyong Jiang, Fan Wang, Jiasi Shen, Sungju Kim, and Sunghun Kim.
\newblock A survey on large language models for code generation.
\newblock {\em arXiv preprint arXiv:2406.00515}, 2024.

\bibitem{kwiatkowski2019natural}
Tom Kwiatkowski, Jennimaria Palomaki, Olivia Redfield, Michael Collins, Ankur Parikh, Chris Alberti, Danielle Epstein, Illia Polosukhin, Jacob Devlin, Kenton Lee, et~al.
\newblock Natural questions: a benchmark for question answering research.
\newblock {\em Transactions of the Association for Computational Linguistics}, 7:453--466, 2019.

\bibitem{vllm}
Woosuk Kwon, Zhuohan Li, Siyuan Zhuang, Ying Sheng, Lianmin Zheng, Cody~Hao Yu, Joseph Gonzalez, Hao Zhang, and Ion Stoica.
\newblock Efficient memory management for large language model serving with pagedattention.
\newblock In {\em Proceedings of the 29th Symposium on Operating Systems Principles}, pages 611--626, 2023.

\bibitem{Laskaridis_2021}
Stefanos Laskaridis, Alexandros Kouris, and Nicholas~D. Lane.
\newblock Adaptive inference through early-exit networks: Design, challenges and directions.
\newblock In {\em Proceedings of the 5th International Workshop on Embedded and Mobile Deep Learning}, MobiSys ’21. ACM, June 2021.

\bibitem{li2023enabling}
Jinhao Li, Shiyao Li, Jiaming Xu, Shan Huang, Yaoxiu Lian, Jun Liu, Yu~Wang, and Guohao Dai.
\newblock Enabling fast 2-bit llm on gpus: Memory alignment, sparse outlier, and asynchronous dequantization.
\newblock {\em arXiv preprint arXiv:2311.16442}, 2023.

\bibitem{li2024largelanguagemodelinference}
Jinhao Li, Jiaming Xu, Shan Huang, Yonghua Chen, Wen Li, Jun Liu, Yaoxiu Lian, Jiayi Pan, Li~Ding, Hao Zhou, Yu~Wang, and Guohao Dai.
\newblock Large language model inference acceleration: A comprehensive hardware perspective, 2024.

\bibitem{li2024eagle}
Yuhui Li, Fangyun Wei, Chao Zhang, and Hongyang Zhang.
\newblock Eagle: Speculative sampling requires rethinking feature uncertainty.
\newblock In {\em International Conference on Machine Learning}, 2024.

\bibitem{lin2024awqactivationawareweightquantization}
Ji~Lin, Jiaming Tang, Haotian Tang, Shang Yang, Wei-Ming Chen, Wei-Chen Wang, Guangxuan Xiao, Xingyu Dang, Chuang Gan, and Song Han.
\newblock Awq: Activation-aware weight quantization for llm compression and acceleration, 2024.

\bibitem{ma2024first}
Chi Ma, Mincong Huang, Ying Zhang, Chao Wang, Yujie Wang, Lei Yu, Chuan Liu, and Wei Lin.
\newblock First activations matter: Training-free methods for dynamic activation in large language models.
\newblock {\em arXiv preprint arXiv:2408.11393}, 2024.

\bibitem{miller1991contextual}
George~A Miller and Walter~G Charles.
\newblock Contextual correlates of semantic similarity.
\newblock {\em Language and cognitive processes}, 6(1):1--28, 1991.

\bibitem{mirzadeh2023relu}
Iman Mirzadeh, Keivan Alizadeh, Sachin Mehta, Carlo~C Del~Mundo, Oncel Tuzel, Golnoosh Samei, Mohammad Rastegari, and Mehrdad Farajtabar.
\newblock Relu strikes back: Exploiting activation sparsity in large language models.
\newblock {\em arXiv preprint arXiv:2310.04564}, 2023.

\bibitem{nallapati2016abstractive}
Ramesh Nallapati, Bowen Zhou, Caglar Gulcehre, Bing Xiang, et~al.
\newblock Abstractive text summarization using sequence-to-sequence rnns and beyond.
\newblock {\em arXiv preprint arXiv:1602.06023}, 2016.

\bibitem{CUTLASS}
NVIDIA.
\newblock Cutlass: Cuda templates for linear algebra subroutines.
\newblock [Online], 2017.
\newblock \url{https://github.com/NVIDIA/cutlass}.

\bibitem{peng2018optimus}
Yanghua Peng, Yixin Bao, Yangrui Chen, Chuan Wu, and Chuanxiong Guo.
\newblock Optimus: an efficient dynamic resource scheduler for deep learning clusters.
\newblock In {\em Proceedings of the Thirteenth EuroSys Conference}, pages 1--14, 2018.

\bibitem{raposo2024mixtureofdepths}
David Raposo, Sam Ritter, Blake Richards, Timothy Lillicrap, Peter~Conway Humphreys, and Adam Santoro.
\newblock Mixture-of-depths: Dynamically allocating compute in transformer-based language models, 2024.

\bibitem{rosenblatt1958perceptron}
Frank Rosenblatt.
\newblock The perceptron: a probabilistic model for information storage and organization in the brain.
\newblock {\em Psychological review}, 65(6):386, 1958.

\bibitem{socher2013recursive}
Richard Socher, Alex Perelygin, Jean Wu, Jason Chuang, Christopher~D Manning, Andrew~Y Ng, and Christopher Potts.
\newblock Recursive deep models for semantic compositionality over a sentiment treebank.
\newblock In {\em Proceedings of the 2013 conference on empirical methods in natural language processing}, pages 1631--1642, 2013.

\bibitem{song2023powerinfer}
Yixin Song, Zeyu Mi, Haotong Xie, and Haibo Chen.
\newblock Powerinfer: Fast large language model serving with a consumer-grade gpu, 2023.

\bibitem{csqa}
Alon Talmor, Jonathan Herzig, Nicholas Lourie, and Jonathan Berant.
\newblock Commonsenseqa: A question answering challenge targeting commonsense knowledge, 2019.

\bibitem{alpaca}
Rohan Taori, Ishaan Gulrajani, Tianyi Zhang, Yann Dubois, Xuechen Li, Carlos Guestrin, Percy Liang, and Tatsunori~B. Hashimoto.
\newblock Stanford alpaca: An instruction-following llama model.
\newblock \url{https://github.com/tatsu-lab/stanford_alpaca}, 2023.

\bibitem{llama}
Hugo Touvron, Louis Martin, Kevin Stone, Peter Albert, Amjad Almahairi, Yasmine Babaei, Nikolay Bashlykov, Soumya Batra, Prajjwal Bhargava, Shruti Bhosale, et~al.
\newblock Llama 2: Open foundation and fine-tuned chat models.
\newblock {\em arXiv preprint arXiv:2307.09288}, 2023.

\bibitem{huggingface}
Thomas Wolf, Lysandre Debut, Victor Sanh, Julien Chaumond, Clement Delangue, Anthony Moi, Pierric Cistac, Tim Rault, Rémi Louf, Morgan Funtowicz, Joe Davison, Sam Shleifer, Patrick von Platen, Clara Ma, Yacine Jernite, Julien Plu, Canwen Xu, Teven~Le Scao, Sylvain Gugger, Mariama Drame, Quentin Lhoest, and Alexander~M. Rush.
\newblock Transformers: State-of-the-art natural language processing.
\newblock In {\em Proceedings of the 2020 Conference on Empirical Methods in Natural Language Processing: System Demonstrations}, pages 38--45, Online, October 2020. Association for Computational Linguistics.

\bibitem{wu2023autogen}
Qingyun Wu, Gagan Bansal, Jieyu Zhang, Yiran Wu, Shaokun Zhang, Erkang Zhu, Beibin Li, Li~Jiang, Xiaoyun Zhang, and Chi Wang.
\newblock Autogen: Enabling next-gen llm applications via multi-agent conversation framework.
\newblock {\em arXiv preprint arXiv:2308.08155}, 2023.

\bibitem{grok}
xAI.
\newblock Open release of grok-1.
\newblock [Online], 2024.
\newblock \url{https://github.com/xai-org/grok-1}.

\bibitem{jiang2024dllm}
yikun jiang, Huanyu Wang, Lei Xie, Hanbin Zhao, Chao Zhang, Hui Qian, and John~C.S. Lui.
\newblock D-{LLM}: A token adaptive computing resource allocation strategy for large language models.
\newblock In {\em The Thirty-eighth Annual Conference on Neural Information Processing Systems}, 2024.

\bibitem{mtbench}
Lianmin Zheng, Wei-Lin Chiang, Ying Sheng, Siyuan Zhuang, Zhanghao Wu, Yonghao Zhuang, Zi~Lin, Zhuohan Li, Dacheng Li, Eric~P. Xing, Hao Zhang, Joseph~E. Gonzalez, and Ion Stoica.
\newblock Judging llm-as-a-judge with mt-bench and chatbot arena, 2023.

\bibitem{zhou2024survey}
Zixuan Zhou, Xuefei Ning, Ke~Hong, Tianyu Fu, Jiaming Xu, Shiyao Li, Yuming Lou, Luning Wang, Zhihang Yuan, Xiuhong Li, et~al.
\newblock A survey on efficient inference for large language models.
\newblock {\em arXiv preprint arXiv:2404.14294}, 2024.

\bibitem{openai_consumption}
Zodhya.
\newblock Optimizing inference on large language models with nvidia tensorrt-llm, now publicly available.
\newblock [Online], 2023.
\newblock \url{https://medium.com/@zodhyatech/how-much-energy-does-chatgpt-consume-4cba1a7aef85}.

\end{thebibliography}

\end{document}